\pdfoutput=1

\documentclass[journal]{IEEEtran}

\usepackage[cmex10]{amsmath}

\usepackage{multirow}
\usepackage{pstricks, pst-plot, pst-grad, pstricks-add, pst-node, pst-3d, pstricks-add}

\usepackage{algorithm}
\usepackage{algpseudocode}
\usepackage{hyperref}

\usepackage{graphicx}
\usepackage{epstopdf}
\usepackage{amssymb}
\usepackage{bm}
\usepackage{amsthm}
\usepackage{array}
\usepackage{enumerate}

\DeclareMathAlphabet      {\mathbfit}{OML}{cmm}{b}{it}
\DeclareMathAlphabet      {\mathbfrm}{OT1}{cmr}{b}{n}

\theoremstyle{plain}% default

\theoremstyle{definition}

\theoremstyle{remark}
\newtheorem{rem}{Remark}

\newcommand{\slfrac}[2]{\left.#1\middle/#2\right.}

\begin{document}
%\linenumbers
%\linenumbersep 3pt\relax


\title{Channel Training for Analog FDD Repeaters:\\ Optimal Estimators and Cram\'{e}r-Rao Bounds}

\author{
\IEEEauthorblockN{Stefan Wesemann and Thomas L. Marzetta}\\
\IEEEauthorblockA{Bell Labs, Nokia\\
Email: \{stefan.wesemann,tom.marzetta\}@nokia-bell-labs.com}

\author{Stefan~Wesemann and
Thomas~L.~Marzetta,~\IEEEmembership{Fellow,~IEEE}}% <-this % stops a space

\thanks{\copyright 2017 IEEE. Personal use of this material is permitted. Permission from IEEE must be obtained for all other uses, in any current or future media, including reprinting/republishing this material for advertising or promotional purposes, creating new collective works, for resale or redistribution to servers or lists, or reuse of any copyrighted component of this work in other works.}
\thanks{S. Wesemann is with Nokia Bell Labs, Stuttgart, Germany, e-mail: stefan.wesemann@nokia-bell-labs.com.} 
\thanks{T.L. Marzetta is with Nokia Bell Labs, Murray Hill, USA, e-mail: tom.marzetta@nokia-bell-labs.com.} %<-this % stops a space
}

%% The paper headers
%\markboth{IEEE TRANSACTIONS ON SIGNAL PROCESSING, accepted for publication}%
%{Distributed Asynchronous Optimization Framework for the MISO Interference Channel}
%% The only time the second header will appear is for the odd numbered pages
%% after the title page when using the twoside option.
%% 
%% *** Note that you probably will NOT want to include the author's ***
%% *** name in the headers of peer review papers.                   ***
%% You can use \ifCLASSOPTIONpeerreview for conditional compilation here if


\maketitle

\begin{abstract}
For frequency division duplex channels, a simple pilot loop-back procedure has been proposed that allows the estimation of the UL \& DL channels at an antenna array without relying on any digital signal processing at the terminal side. For this scheme, we derive the maximum likelihood (ML) estimators for the UL \& DL channel subspaces, formulate the corresponding Cram\'{e}r-Rao bounds and show the asymptotic efficiency of both (SVD-based) estimators by means of Monte Carlo simulations. In addition, we illustrate how to compute the underlying (rank-1) SVD with quadratic time complexity by employing the power iteration method. To enable power control for the data transmission, knowledge of the channel gains is needed. Assuming that the UL \& DL channels have on average the same gain, we formulate the ML estimator for the channel norm, and illustrate its robustness against strong noise by means of simulations.
\end{abstract}

\begin{IEEEkeywords}
Analog feedback, subspace estimation, fronthaul channel, frequency-division duplex, analog repeater, channel state information (CSI), multiple-input single-output (MISO)
\end{IEEEkeywords}

\IEEEpeerreviewmaketitle

% ##########################################################################
\section{Introduction}\label{sec:intro}

\IEEEPARstart{C}{hannel}-state information (CSI) plays a key role in a multi-user MIMO system. With CSI, the antenna array can send multiple messages, simultaneously and selectively, to autonomous terminals. Time-division duplex (TDD) systems offer a straightforward way for the antenna array to acquire the CSI. Since the downlink (DL) and uplink (UL) channels in a TDD system share the same frequency, reciprocity implies that the antenna array can learn the DL channel from known pilot signals on the UL. For frequency-division duplex (FDD) systems, an additional DL CSI estimation and feedback step is needed, which requires digital signal processing at the terminal side. The terminals first estimate the DL channel from pilot signals transmitted by the antenna array, and then they feed back the CSI to the array on the known UL channel. 

In \cite{fcPatent}, an analog feedback based channel training scheme has been proposed for FDD systems, which requires no digital processing capabilities at the terminals. The key idea is to employ analog repeaters for the channel estimation, which simply retransmit (i.e., amplify, mix and forward) the received DL pilot signal on the UL channel. Given the received UL signal, the antenna array has to estimate both the UL \& DL channels. In order to separate the CSI feedback from multiple repeaters, the channel training for the individual repeaters is performed in time multiplex; that is, each repeater is trained in a dedicated time slot. The envisaged use case for this training scheme is outlined in Section \ref{sec:usecase}. The present paper focuses on the analysis of the underlying estimation problem.

As described in Section \ref{sec:prob_form}, the received signal on a particular subband contains the (outer) product of the unknown UL \& DL channel vectors, so that each channel vector can be estimated only up to an unknown scalar scaling factor. Consequently, the only observable invariants of the propagation channels are their subspaces. Therefore, we formulate in Section \ref{sec:ss_est} the maximum likelihood (ML) estimators for the UL \& DL subspaces. Subject to some familiar simplifying assumptions [independent identically distributed (i.i.d.) channel coefficients and noise components] that serve as a viable approximation for non-line-of-sight channels below 6 GHz, we obtain closed-form solutions for the ML estimators. Moreover, based on the intrinsic Cram\'{e}r-Rao bounds (CRB) derived in \cite{1420804}, we are able to formulate bounds on the achievable subspace estimation accuracy. In Section \ref{compl_reduc}, we discuss methods that allow a significant reduction of the subspace estimator's computational complexity, and evaluate these by means of Monte Carlo simulations in Section \ref{sec:monte_carlo}. Moreover, assuming knowledge about the repeater's average transmit power and the variance of the UL perturbations at the antenna array, we formulate in Section \ref{sec:ulChanNormEst} the ML estimator for UL channel vector norm, which is needed for the power control of the data streams. Finally, we summarize our findings and conclude with a discussion of open problems in Section \ref{sec:sum_disc}.

To encourage reproducibility and extensions to this paper, all the simulation results can be generated by the R code \cite{rcite} that is available at \url{https://github.com/stefanwesemann/Channel-Training-Analog-FDD-Repeater}.

\begin{figure*}[t]
\centering
\includegraphics[width=16.0cm]{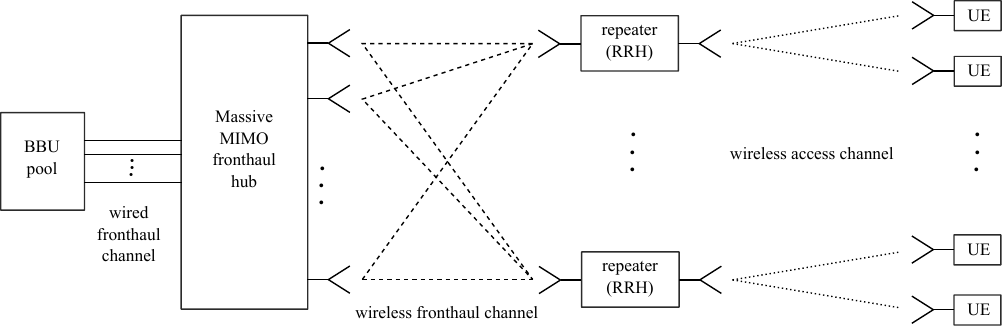}
\caption{The envisaged use case: A C-RAN architecture with a wireless fronthaul, established by a massive MIMO fronthaul hub and a set of analog repeaters. The latter replace the regular remote radio heads (RRHs), and can be fully energy-autonomous (e.g., driven by a battery that is charged by a small size solar panel). The fronthaul and access channels (dashed and dotted lines, respectively) use dedicated frequency bands, each hosting paired UL \& DL carriers for FDD operation.
The wireless fronthaul (i.e., the cascading of fronthaul hub, channel and repeater) replaces a regular wired fronthaul network (e.g., CPRI-fibers), and is completely transparent to the baseband units (BBUs) and user equipment (UE). As illustrated, each repeater (RRH) represents a single-antenna small cell, which serves its associated users in time or frequency division multiple access.}
\label{fig:ar_deploy}
\end{figure*}

\subsection{Use Case: Wireless Fronthauling for Cloud-RAN Systems}\label{sec:usecase}
In the Cloud Radio Access Network (C-RAN) architecture \cite{6897914}, the baseband processing is centralized and shared among sites in a virtualized baseband unit (BBU) pool. This allows a more efficient resource utilization (e.g., by sharing between different network operators), lower delays for network coordination and joint processing schemes (e.g., enhanced ICIC and CoMP in LTE-Advanced), improved scalability and ease of network maintenance.
The BBUs are connected to remote radio heads (RRHs) through low latency, high bandwidth fronthaul links such as fiber (e.g., CPRI, OBSAI), free space optics or point-to-point millimeter wave (mmWave). However, such static fronthaul links rely on the availability of fiber or have to cope with propagation challenges and incurred resilience issues.

\begin{figure*}[t]
\centering
\includegraphics[width=18.0cm]{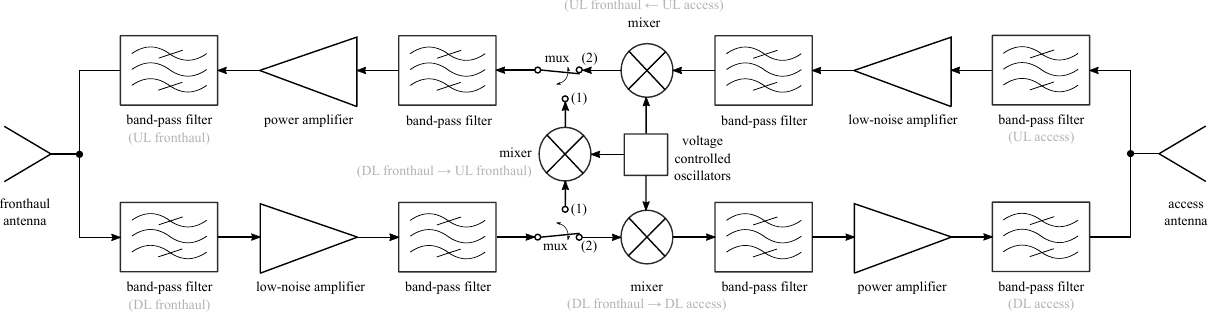}
\caption{Block diagram of an analog repeater architecture, which filters and amplifies the received signal from the DL fronthaul frequency (resp. UL access frequency), and converts this signal to the DL access frequency (resp. UL fronthaul frequency). The two multiplexers (mux) switch simultaneously between two modes: (1) \emph{fronthaul channel training} where the received pilot signal from the DL fronthaul is retransmitted on the UL fronthaul, (2) \emph{signal forwarding} from fronthaul to access and vice versa. The multiplexers can be triggered by using e.g. a dedicated chirp signal.}
\label{fig:ar_schematic}
\end{figure*}

A promising alternative is massive MIMO \cite{5595728} based wireless fronthauling. At carrier frequencies below 6 GHz, massive MIMO enables adaptive and cost-efficient point-to-multi-point connections even under non-line-of-sight conditions.
As illustrated in Figure \ref{fig:ar_deploy}, the BBU-pool is connected to a massive MIMO fronthaul hub (FH), which multiplexes the IQ-data streams for the individual RRHs\footnote{In 4G LTE, a typical RRH has at least two antenna ports, each driven by a dedicated IQ-data stream. In the described fronthaul architecture, each RRH antenna port can be represented by a dedicate repeater unit.} over the same time/frequency resources, thereby realizing multiple ``wireless CPRI''-links. Assuming dedicated frequency bands for the fronthaul and access channels, the RRH architecture degenerates to a simple repeater\footnote{A repeater is a physical layer (i.e., layer 1 in the OSI model) relay equipment that amplifies e.g., DL signals from a base station for transmission to a terminal. It is an Amplifier and Forward (AF) type of relay \cite{1362898}, in contrast to a layer 2 relay which is a Decode and Forward (DF) type of relay. The demodulation and decoding processing performed at the layer 2 relay overcomes the drawback in layer 1 relays of deteriorated received SINR caused by amplification of inter-cell interference and noise. At the same time, however, the layer 2 relay causes a delay associated with modulation/demodulation and encoding/decoding processing, and requires advanced radio functions such as mobility control and retransmission control by automatic repeat request.} structure which translates the signals between the fronthaul and access bands (i.e., no full-duplex operation \cite{5161790} is needed). By focusing on FDD systems (i.e., fronthaul and access band are each split into an UL \& DL subband, separated by a frequency offset), such a repeater can be realized with pure analog components because the UL \& DL channels are separated by bandpass filters (i.e., no transmit-receive switching is required). The resulting repeater architecture is illustrated in Figure \ref{fig:ar_schematic}, and is supposed to have a total power consumption of less than 15 Watts\footnote{In the following, we provide a conservative power budget for the analog repeater architecture as shown in Figure \ref{fig:ar_schematic}. Assuming a minimum average receive signal power of -100 dBm, an average transmit power of 23 dBm and 2 dB loss per band-pass filter (yielding 8 dB total loss per signal path), the low-noise and power amplifier combination has to realize a gain of 131 dB. This can be implemented by using e.g., discrete components such as one ANADIGICS AWB7221 power amplifier (31dB gain, 3 Watts power consumption), and five cascaded NXP BGU8052 low-noise amplifiers (each with 20dB gain, 0.3 Watts power consumption). By assuming 0.53 Watts per voltage controlled oscillator (e.g., Analog Devices ADF4351), the total power consumption of the analog repeater will be 10.6 Watts.
The overall power consumption can be further reduced by using integrated transceiver chains such as used in terminal chip-sets.}. This enables the deployment of energy-autonomous repeaters that are driven by a battery, which is charged by e.g., a small size solar panel.

In order to multiplex several high data-rate\footnote{For the fronthauling of LTE signals, a link SINR of 25 dB with an outage probability of 10 \% should be guaranteed.} fronthaul links over the same time/frequency resource, the FH requires accurate CSI for both, the UL \& DL fronthaul channels. 
Employing the described analog repeater architecture, classical channel training methods for FDD MIMO systems such as \cite{1608543}, \cite{4527202}, \cite{1658228}, \cite{6777295}, \cite{6816089} are not applicable because they rely on digital signal processing capabilities at the repeater side.
Therefore, the analog feedback-based training method was proposed in \cite{fcPatent}. As described in the previous section, the FH transmits pilots in the DL fronthaul, which are directly fed back by a single repeater on the UL fronthaul.
The increased training overhead, which scales with the number of FH antennas and the number of repeaters (due to their sequential training), is compensated by the large coherence block size of fixed wireless channels. Channel measurements (see e.g., \cite[Table I]{08123235}) have shown channel coherence times in the order of 50 ms to 100 ms. Assuming a 3GPP Urban Micro channel with a root mean square delay spread of 0.25 $\mu$s, a suitable subband size for channel training is 180 kHz. By adopting the LTE numerology, this corresponds to a coherence block size of at least 50 ms $\times$ 12 symbols/subband $\times $ 14 symbols/ms = 8400 symbols. In comparison to this large coherence block size, the channel training for a FH with e.g., 64 antennas and 8 repeaters requires 64 symbols $\times$ 8 = 512 symbols, which make up less than 10 \% overhead.

One should note that the transmission of the UL \& DL IQ-data streams in the fronthaul must be suspended during the channel training. LTE offers a simple mean for creating such transmission gaps in its DL stream, namely by inserting almost empty subframes known as MBSFN (Multimedia Broadcast Single Frequency Network) subframes \cite{LTEbook}. For creating empty slots in the LTE UL, however, special scheduler tweaks are required. Given that mode of operation, the entire wireless fronthaul (including the fronthaul channel training) will be fully transparent to the BBUs and terminals. In the access channel, the analog repeaters will be indistinguishable from regular RRHs. On the fronthaul channel, the FH mimics a set of regular RRHs to the BBUs.
The feasibility of such a system is currently investigated within Bell Labs' F-Cell project. The project target is to support at least 8 repeaters with 64 antennas at the FH.

\subsection{Related Work}
The utility of \emph{analog CSI transfer} is suggested by the known optimality of uncoded analog linear modulation when a white Gaussian random message process is transmitted over an additive Gaussian white noise channel, subject to a minimum mean-square error criterion \cite{1053821}, \cite{1197846}. The concept of analog linear modulation has been adopted for MIMO systems in e.g., \cite{1608543}, \cite{4527202}, \cite{1658228}. In \cite{1608543}, the relative effort in TDD and FDD systems that is needed for the antenna array to acquire the same-quality UL \& DL CSI has been investigated. The authors assume perfect DL CSI at the terminals, and the use of analog linear modulation for the DL CSI feedback on the UL channel. They show that the multi-antenna effect apparently benefits CSI transfer in the UL channel. As the number of antennas increases, the FDD CSI feedback burden decreases despite the fact that the users have to transfer increasing amounts of CSI. Thus, adding redundancy to the feedback signal such as illustrated in \cite{4100513} is only necessary for a very small number of array antennas. The practical feasibility of the analog feedback scheme has been investigated in \cite{6399275} for a small $2 \times 2$ multi-user MIMO system.

The formulation of the maximum likelihood (ML) \emph{subspace estimator} for an UL channel has been described in e.g., \cite{1420804}. For the DL subspace estimation, however, this result is not directly applicable because of the presence of complex double Gaussian terms in our system model. By a step-wise formulation of the ML estimators along a Markov chain, we are able derive a closed-form solution for the DL ML estimator. Of course, for more complicated settings one has to resort to e.g., Newton or conjugate gradient algorithms, whose formulation on the Grassmann and Stiefel manifolds is provided in \cite{edelman1998geometry}. The UL subspace estimation in a Bayesian setting has been described in \cite{839985}, \cite{6006540}. In \cite{6006540}, the authors formulate the optimal subspace estimator which minimizes the mean square (subspace) distance. By adopting the Bingham distribution as a prior, a closed-form estimator is obtained. We briefly outline this approach in Section \ref{sec:bayesian}.
For the derivation of the Cram\'{e}r-Rao bounds (CRBs) for our UL \& DL subspace estimators, we again build upon the results from \cite{1420804}, which provides intrinsic CRBs for estimation problems on arbitrary manifolds. A brief description of the herein used Grassmann manifold is given in Section \ref{prelim}. 
The application of subspace estimation techniques to TDD MIMO channels has been proposed in \cite{1323251}, \cite{6288608}, \cite{7439748}, \cite{6666355}.
In \cite{1323251}, an iterative estimation scheme for a TDD single-user MIMO system is described. Assuming that channel reciprocity holds, the antenna array and terminal perform power iterations over-the-air; that is, they send their singular vector estimates back and forth until convergence. However, the convergence of this algorithm is shown numerically only, and there exists no lower bound on the achievable estimation error. The extension of this training method to hybrid transceiver architectures (employing analog \& digital beamforming) for millimeter wave systems is described in \cite{7439748}.
An eigenvalue decomposition (EVD) based UL channel estimator for a multi-user massive MIMO system has been described in \cite{6288608}. By exploiting the fact that the channel vectors of different terminals are asymptotic orthogonal if the number of array antennas goes to infinity, the subspaces of the channel vectors can be derived from the eigenvectors of the received signal's sample covariance matrix (SCM). However, the proper correspondence between eigenvectors and terminals depends on the a priori knowledge of the distinct long-term path gains. In order to resolve amplitude and phase ambiguities that exists between the estimated subspaces and the channel vectors, the transmission of short UL training sequences is needed. An improved EVD-based estimation (using a widely-linear algorithm) as well as closed-form expressions for the mean-square channel estimation error are provided in \cite{6666355}.

The application of \emph{massive MIMO for fronthauling} is a viable option for 5G communication systems, see e.g., \cite{6894457}, \cite[Section 12.4]{3924}. Current research primarily focuses on its application in the mmWave range. The use of phased antenna arrays with electronic beam-steering reduces installation cost, provides beam-tracking and interference management. Such wireless fronthaul links typically rely on line-of-sight-dominant channels and are limited to distances less than 1 km. These limitations can be resolved when using frequency bands below 6 GHz, but at the cost of complex hardware architectures (i.e., a larger number of digital transceiver chains) and advanced CSI acquisition methods.
For FDD systems, however,  DL training and feedback overhead scales with the number of array antennas and can be quite overwhelming. Overhead reduction methods have been proposed in e.g., \cite{6777295}, \cite{6816089}, and typically assume either some form of channel sparsity that can be utilized, or some a priori channel knowledge such as long-term channel statistics.
In \cite{6777295}, the spatial and temporal channel correlations are exploited for the careful design of training pilots with a reduced length. The (hidden) joint sparsity structure in the user channel matrices is utilized in \cite{6816089} by means of a distributed, compressive sensing (CS) technique. 
One should note that the analog DL CSI feedback concept can be adopted for both training methods, but at expense of an additional UL channel training step.
Methods that focus on the reduction of the DL CSI feedback (i.e., feedback compression) can be found in e.g., \cite{7063458} (projection based), \cite{6214417} (CS based), \cite{6884026} (pattern based). The application of these techniques to analog feedback schemes is proposed in \cite{6881223} for feedback reduction, and in \cite{7302003}, \cite{7179337} to improve the CSI estimation using CS techniques. The application of these overhead reduction methods to the described wireless fronthaul use case is subject to future work.

\subsection{Notation and Preliminaries} \label{prelim}
Vectors and matrices are given in lowercase and uppercase boldface letters, respectively. The $m$-th component of the vector $\mathbfit{y}$ is denoted by $\mathbfit{y}_{[m]}$, the $n$-th column of the matrix $\mathbfit{Y}$ is denoted by $\mathbfit{Y}_{[:,n]}$. $(\cdot)^H$ denotes the Hermitian transpose. We use the symbol $\operatorname{E}$ for the expectation operator. $\boldsymbol 1_M$ denotes the all-ones $M\times 1$ vector, $\mathbfit{I}_M$ denotes the $M\times M$ identity matrix. The function $\mathrm{exp}\{\cdot\}$ stands for the exponential, $\mathrm{etr}\{\cdot\}$ stands for the exponential of the trace of the matrix between the braces, and $\delta(\cdot)$ represents the Dirac delta function. By the dominant eigenvector $\mathbfit{v}_\mathrm{max}(\mathbfit{Y})$ of a matrix ${\mathbfit{Y}}$, we denote the eigenvector that corresponds to the largest (i.e., dominant) eigenvalue $\lambda_\mathrm{max}(\mathbfit{Y})$ of ${\mathbfit{Y}}$. Finally, $\Gamma(x)=\int_0^\infty \lambda^{x-1}\mathrm{exp}\left\{-\lambda\right\}d\lambda$ is the gamma function.

Two (non-zero) vectors $\mathbfit{x},\mathbfit{y}\in\mathbb{C}^M$ are called equivalent $\mathbfit{x}\sim \mathbfit{y}$ in terms of subspaces if and only if there exists a (non-zero) scalar $a\in\mathbb{C}$ such that $\mathbfit{x}=a\cdot\mathbfit{y}$. This relation groups the vectors in $\mathbb{C}^M$ into equivalence classes. The equivalence class of a vector $\mathbfit{x}\in\mathbb{C}^M$ is denoted by $[\mathbfit{x}]$, and is defined as the set $[\mathbfit{x}]=\left\{\mathbfit{y}\in\mathbb{C}^M|\mathbfit{y}\sim \mathbfit{x}\right\}$. The set of all equivalence classes forms the Grassmannian $G_{1}(\mathbb{C}^M)$; that is, the set of one-dimensional subspaces in $\mathbb{C}^M$. More generally, we denote the set of $p$-dimensional subspaces in $\mathbb{C}^M$ by $G_{p}(\mathbb{C}^M)$. A closely related manifold is the Stiefel manifold $V_p(\mathbb{C}^M)$, which represents the set of $M\times p$ (complex) matrices with orthonormal columns. While each point in the Stiefel manifold $V_p(\mathbb{C}^M)$ is uniquely represented by a $M\times p$ matrix, the Grassmannian $G_{p}(\mathbb{C}^M)$ consists of subspaces which may be specified by an arbitrary orthogonal basis stored as an $M\times p$ matrix.
One should note that $V_1(\mathbb{C}^M)$ is equivalent to the $M-1$ dimensional complex hypersphere $\mathbb{S}^{M-1} \subset \mathbb{C}^M$.

% ##########################################################################

\section{Problem Formulation} \label{sec:prob_form}

\subsection{System Model and Pilot Transmission}
As discussed in Section \ref{sec:intro}, we consider a time-invariant, frequency-flat FDD system, which consists of a antenna array (AA) with $M$ antennas and a single-antenna repeater. 
The UL \& DL propagation channels, denoted by $\mathbfit{g}$ and $\mathbfit{h}$, respectively, are both modeled as $M\times 1$ complex vectors whose elements are i.i.d. $\mathcal{CN}(0,\beta)$. 

In the DL, the AA antennas transmit orthogonal pilot signals of length\footnote{For scenarios with limited transmit power and very low signal-to-interference-plus-noise ratios, it is beneficial to choose $\tau > M$. The subsequent pilot reverse modulation step in Section \ref{sec:svdSSEst} provides a noise plus interference suppression. Of course, in a massive MIMO setting choosing $\tau > M$ might be to costly in terms of required training overhead.} $\tau\geq M$, which are defined by the columns of the $\tau\times M $ unitary matrix $\boldsymbol{\Phi}$, where $\boldsymbol{\Phi}^H\boldsymbol{\Phi} = \mathbfit{I}_M$. Assuming an average transmit power per antenna of $P/M$, the repeater receives the weighted superposition of all pilot signals, 
\begin{align}
\mathbfit{x}^H = \sqrt{\tau/ MP} \mathbfit{h}^H \boldsymbol{\Phi}^H + \mathbfit{w}^H, \label{eq:dl_rcvd_sgnl}
\end{align}
perturbed by the $\tau \times 1$ vector $\mathbfit{w}$ which represents the receiver noise and DL interference. We model the elements of $\mathbfit{w}$ as i.i.d. complex Gaussian noise, and take the noise variance to be 1 to minimize notation. The DL signal-to-interference-plus-noise ratio (SINR) $\rho_\mathrm{D}$, which is defined as the ratio of the average receive power to the additive noise plus interference variance at the repeater, assuming that all AA transmit power is fed into a single antenna\footnote{By using this single-antenna equivalent transmit power, we obtain a SINR definition that is independent from any beamformer applied by the AA, see e.g., \cite[Section 2.2.1]{marzetta_larsson_yang_ngo_2016}.}, is given by
\begin{align}
\rho_\mathrm{D} = \beta P/1 = \beta P.
\end{align}

The repeater uses analog linear modulation \cite{1608543} in order to feed back the received pilot signal to the AA. In this process, the repeater translates the DL carrier frequency to the UL carrier frequency, and scales the average power of the feedback signal by factor $\alpha$. The resulting average transmit power of the repeater is given by $Q=\alpha(\beta P +1)$ where the ``1'' represents the normalized DL noise variance.
In the UL, the AA receives $M$ replicas of the DL feedback signal, which are individually scaled by the components of the channel vector $\mathbfit{g}$, and perturbed by the receiver noise and UL interference $\mathbfit{N}$, whose elements are i.i.d. complex Gaussian with unit variance. The $M \times \tau$ receive signal is 
\begin{align}
\mathbfit{Y} &= \sqrt{\alpha} \mathbfit{g} \mathbfit{x}^H + \mathbfit{N}. \label{eq:ul_rcvd_sgnl}
\end{align}
The UL SINR $\rho_\mathrm{U}$, which we define as the ratio of the average receive power to the additive perturbation variance at an arbitrary antenna of the array, is given by
\begin{align}
\rho_\mathrm{U} = \beta Q/1 = \alpha\beta(\beta P+1).
\end{align}

\subsection{Pilot Reverse Modulation and Observable Invariants}\label{sec:svdSSEst}
Based on the received signal $\mathbfit{Y}$, the AA has to estimate the UL \& DL channels. The first processing step is the pilot reverse modulation, which consists of the right-multiplication of \eqref{eq:ul_rcvd_sgnl} by $\boldsymbol{\Phi}$. The resulting $M \times M$ observation matrix is
\begin{align}
\tilde{\mathbfit{Y}} &= \mathbfit{Y}\boldsymbol{\Phi} \nonumber\\
&=\sqrt{\alpha} \mathbfit{g} \mathbfit{x}^H\boldsymbol{\Phi} + \mathbfit{N}\boldsymbol{\Phi}\nonumber\\
&=\sqrt{\alpha\tau/M P}\mathbfit{g} \mathbfit{h}^H+\sqrt{\alpha}\mathbfit{g}\tilde{\mathbfit{w}}^H + \tilde{\mathbfit{N}}. \label{effSysModel}
\end{align}
The elements of the transformed $M\times 1$ DL perturbation $\tilde{\mathbfit{w}}$ and the $M\times M$ UL perturbation $\tilde{\mathbfit{N}}$ remain i.i.d. complex Gaussian with unit variance. The pilot reverse modulation can be interpreted as a matched filtering of the received signal with the transmitted one. For the case $\tau>M$, this operation yields a noise plus interference suppression, as reflected by the post-processing UL \& DL SINRs,
\begin{align}
\tilde{\rho}_\mathrm{D}&=\tau/M\beta P,\\
\tilde{\rho}_\mathrm{U}&=\alpha\beta(\tau/M \beta P  + 1).
\end{align}
One should note that these SINRs apply only for the channel training period.

An inspection of \eqref{effSysModel} reveals that the UL \& DL channel vectors cannot be estimated explicitly because each vector is only observable as the outer product with an unknown vector.
The only \emph{identifiable} invariants of both, the UL \& DL channels, are the subspaces $[\mathbfit{g}]$ and $[\mathbfit{h}]$, respectively, which are uniquely represented as elements of the Grassmannian. In addition, given the knowledge of average repeater power $Q$ and the variance of the UL perturbations, one can estimate the (squared) norm of the UL channel vector.
Note that this squared norm serves also as a reasonable estimate for DL channel vector norm because on average the DL frequency response should have approximately the same power as the UL frequency response.
Thus, the only residual uncertainty in the acquired CSI is a complex phase rotation. However, it will not affect the fronthaul operations, as the internal pilots carried by the e.g., LTE signals will automatically compensate for any phase offsets.\\

% ##########################################################################

\section{Maximum Likelihood Subspace Estimation} \label{sec:ss_est}

\subsection{Uplink Subspace Estimation and CRB}\label{UlSec}
In the following, we associate elements in the Grassmannian $G_1(\mathbb{C}^M)$ with points on the hypersphere $\mathbb{S}^{M-1}$, which provides a simple problem parametrization based on Cartesian coordinates. One should note that the map between these two manifolds is not a bijection as two antipodal points on the hypersphere correspond to the same subspace. However, this ambiguity is absorbed by the likelihood function that we want to maximize in the following.

Without loss of generality, we can split the UL channel vector $\mathbfit{g}$ into two independent random variables, 
\begin{align}
\mathbfit{g}=\sqrt{\zeta}\boldsymbol{\phi},\label{eq:ulVecSplit}
\end{align}
where $\zeta=\left\|\mathbfit{g}\right\|^2$ is $\slfrac{1}{2}$ times a chi-square random variable with $2M$ degrees of freedom and mean value $\beta M$, and $\boldsymbol{\phi}=1/\left\|\mathbfit{g}\right\|\mathbfit{g}$ is an isotropically distributed (i.d.) random unit vector. Their probability distribution functions (pdfs) are given by
\begin{align}
p(\zeta) = \frac{1}{\beta \Gamma(M)} \left({\zeta}/{\beta}\right)^M \mathrm{exp} \left\{-{\zeta}/{\beta}\right\}
\end{align}
and
\begin{align}
p(\boldsymbol\phi) = \frac{\Gamma(M)}{\pi^M}\delta(\boldsymbol\phi^H \boldsymbol\phi - 1).\label{eq:pdfPhi}
\end{align}
Based on the observation matrix $\tilde{\mathbfit{Y}}$, the ML estimate for the UL subspace (i.e., represented by a point on $\mathbb{S}^{M-1}$) is obtained as
\begin{align}
\hat{\mathbfit{g}}^\mathrm{ML} &\sim \mathrm{arg}\;\underset{\boldsymbol\phi\in\mathbb{S}^{M-1}}{\mathrm{max}}\; p(\tilde{\mathbfit{Y}}|\boldsymbol\phi),\label{eq:argMaxUL}
\end{align}
where the conditional pdf $p(\tilde{\mathbfit{Y}}|\boldsymbol\phi)$ can be written as a marginal pdf,
\begin{align}
p(\tilde{\mathbfit{Y}}|\boldsymbol\phi) = \int p(\tilde{\mathbfit{Y}}|\boldsymbol\phi,\zeta) p(\zeta) d\zeta.
\end{align}
As shown at the end of this section (Eq. \eqref{eq:ul_ML_est}), the ML solution of \eqref{eq:argMaxUL} is independent of $\zeta$. Consequently, we can set
\begin{align}
p(\tilde{\mathbfit{Y}}|\boldsymbol\phi) \propto  p(\tilde{\mathbfit{Y}}|\boldsymbol\phi,\zeta_0),\label{eq:zeta0Prop}
\end{align}
for some arbitrary but fixed $\zeta_0$.

Since $\tilde{\mathbfit{Y}}$ given $\boldsymbol\phi,\zeta_0$ is the sum of two zero-mean random Gaussian matrices, its conditional pdf is
\begin{align}
p(\tilde{\mathbfit{Y}}|\boldsymbol\phi,\zeta_0)=&\left(\pi^M \mathrm{det} \mathbfit{R}_{\tilde{\mathbfit{Y}}|\boldsymbol\phi,\zeta_0} \right)^{-M} \times \nonumber\\
& \mathrm{etr}\left\{-\mathbfit{R}_{\tilde{\mathbfit{Y}}|\boldsymbol\phi,\zeta_0}^{-1} \tilde{\mathbfit{Y}}\tilde{\mathbfit{Y}}^H   \right\}, \label{eq:pdfUL}
\end{align}
where $\mathbfit{R}_{\tilde{\mathbfit{Y}}|\boldsymbol\phi,\zeta_0}$ denotes the conditional covariance matrix for the individual columns of $\tilde{\mathbfit{Y}}$; that is,
\begin{align}
\mathbfit{R}_{\tilde{\mathbfit{Y}}|\boldsymbol\phi,\zeta_0}&= \mathrm{E}\left[\tilde{\mathbfit{Y}}_{[:,n]}\tilde{\mathbfit{Y}}_{[:,n]}^H|\boldsymbol\phi,\zeta_0\right]  \\
&= \alpha \tau/M P \zeta_0 \boldsymbol\phi \mathrm{E}\left[\mathbfit{h}_{[n]}^* \mathbfit{h}_{[n]} \right] \boldsymbol\phi^H  + \nonumber\\
& \quad \quad \alpha  \zeta_0 \boldsymbol\phi \mathrm{E}\left[\tilde{\mathbfit{w}}_{[n]}^* \tilde{\mathbfit{w}}_{[n]} \right] \boldsymbol\phi^H+  \mathrm{E}\left[\tilde{\mathbfit{N}}_{[:,n]} \tilde{\mathbfit{N}}_{[:,n]}^H \right] \nonumber\\
&= \nu \boldsymbol\phi\boldsymbol\phi^H  + \mathbfit{I}_M, \label{eq:ulCovMat}
\end{align}
with $\nu=\alpha\zeta_0(\tau/M\beta P +1 )$ for some arbitrary $n$, $1 \leq n \leq M$.
The computation of its inverse and its determinant yields
\begin{align}
\mathbfit{R}_{\tilde{\mathbfit{Y}}|\boldsymbol\phi,\zeta_0}^{-1}&= \mathbfit{I}_M - \frac{\nu}{\nu +1}\boldsymbol\phi\boldsymbol\phi^H,\label{eq:invR}\\
\mathrm{det}\mathbfit{R}_{\tilde{\mathbfit{Y}}|\boldsymbol\phi,\zeta_0}&=1+\nu\label{eq:detR}.
\end{align}
Since $\tilde{\mathbfit{Y}}$, $\zeta_0$ and thus $\nu$ are fixed, we have the following proportionality
\begin{align}
p(\tilde{\mathbfit{Y}}|\boldsymbol\phi,\zeta_0) &\propto \mathrm{etr}\left\{-\mathbfit{R}_{\tilde{\mathbfit{Y}}|\boldsymbol\phi,\zeta_0}^{-1} \tilde{\mathbfit{Y}}\tilde{\mathbfit{Y}}^H   \right\}\nonumber\\
&\propto  \mathrm{etr}\left\{\frac{\nu}{\nu +1} \boldsymbol\phi^H \tilde{\mathbfit{Y}}\tilde{\mathbfit{Y}}^H \boldsymbol\phi  \right\} \mathrm{etr}\left\{\tilde{\mathbfit{Y}}\tilde{\mathbfit{Y}}^H \right\}^{-1}\nonumber\\
&\propto \mathrm{exp}\left\{\boldsymbol\phi^H \tilde{\mathbfit{Y}}\tilde{\mathbfit{Y}}^H  \boldsymbol\phi\right\}.\label{eq:pYPhiZeta}
\end{align}
One should note that this proportionality is independent from $\zeta_0$, which proves step \eqref{eq:zeta0Prop}. Moreover, it turns out that the likelihood function in \eqref{eq:argMaxUL} is proportional to the complex Bingham distribution \cite{complBingham} on $\mathbb{S}^{M-1}$, with the parameter matrix $\tilde{\mathbfit{Y}}\tilde{\mathbfit{Y}}^H$. For positive semi-definite parameter matrices, \eqref{eq:pYPhiZeta} has two antipodal maxima, which are achieved by the dominant eigenvector of $\tilde{\mathbfit{Y}}\tilde{\mathbfit{Y}}^H$ (denoted by $\mathbfit{v}_\mathrm{max}(\tilde{\mathbfit{Y}}\tilde{\mathbfit{Y}}^H)$), yielding the ML subspace estimate
\begin{align}
\hat{\mathbfit{g}}^\mathrm{ML} \sim \mathbfit{v}_\mathrm{max}(\tilde{\mathbfit{Y}}\tilde{\mathbfit{Y}}^H). \label{eq:ul_ML_est}
\end{align}
\\

A lower bound on subspace estimation accuracy is established by the Cram\'{e}r-Rao bound (CRB), which states that the variance of any unbiased estimator is at least as high as the inverse of the Fisher information metric (FIM). Under certain regularity conditions, the FIM can be expressed using the second derivatives of the log-likelihood function $\log p(\tilde{\mathbfit{Y}}|[\mathbfit{g}])$ with respect to the \emph{subspace} $[\mathbfit{g}]$. For a discussion of the difficulties that arise when computing the FIM on arbitrary manifolds in which no set of intrinsic coordinates exists, we refer the interested reader to \cite{1420804}. In the following, we simply introduce the definitions that are needed for a proper interpretation of the subsequently stated CRBs.

The intrinsic version of the CRB for any unbiased estimator on the Grassmannian $G_p(\mathbb{C}^M)$ as formulated in \cite{1420804}, relies on the natural Riemannian metric of $G_p(\mathbb{C}^M)$. Subspace distances corresponding to this metric are given by the 2-norm of the vector of principal angles between two subspaces. By representing one-dimensional subspaces as points on the hypersphere, the distance between $\boldsymbol\phi,\hat{\boldsymbol\phi}\in\mathbb{S}^{M-1}$ is given by
\begin{align}
d_\mathrm{ss}(\boldsymbol\phi,\hat{\boldsymbol\phi}) = \mathrm{arccos}\left(|\boldsymbol\phi^H\hat{\boldsymbol\phi}|\right). \label{eq:dss}
\end{align}
Based on this distance measure, the mean-square error (MSE) between a fixed $\boldsymbol\phi$ and its estimate $\hat{\boldsymbol\phi}$ is defined as
\begin{align}
\epsilon_{\hat{\boldsymbol\phi}}^2 = \mathrm{E}\left[d_\mathrm{ss}(\boldsymbol\phi,\hat{\boldsymbol\phi})^2\right], \label{eq:mse_def}
\end{align}
where the expectation is with respect to the sampling distribution of $\hat{\boldsymbol\phi}$.
For system models with the form of \eqref{eq:ul_rcvd_sgnl} (i.e., one-dimensional subspaces), \cite{1420804} provides the inverse FIM 
\begin{align}
\mathcal{I}^{-1}(M,T,\gamma) = \frac{(M-1)(1+\gamma)} {T\gamma^2}, \label{eq:inverseFIM}
\end{align}
which is parametrized by 
\begin{description}
	\item[$M$]: Number of array antennas,
	\item[$T$]: Number of i.i.d. observation vectors (here: number of columns of $\tilde{\mathbfit{Y}}$; i.e., $T=M$),
	\item[$\gamma$]: Effective SINR $\gamma=\zeta/\beta\tilde{\rho}_\mathrm{U}$, assuming a unit norm channel vector.
\end{description}
This inverse FIM provides the CRB on the natural subspace distance between the true subspace and any unbiased estimate of it. For a \emph{fixed} realization $\mathbfit{g}$, we have the uplink CRB 
\begin{align}
\epsilon_{\hat{\mathbfit{g}}}^2 \geq \mathcal{I}^{-1}(M,M,\zeta/\beta\tilde{\rho}_\mathrm{U})  \quad (\mathrm{rad}^2). \label{eq:ul_crb_fixedG}
\end{align}
In order to obtain an average CRB for random $\mathbfit{g}$, we have to take the expectation with respect to $\zeta$. By the Jensen's inequality and the convexity of $\mathcal{I}^{-1}(M,T,\gamma)$ with respect to $\gamma$, we obtain the inequality
\begin{align}
\bar{\epsilon}_{\hat{\mathbfit{g}}}^2=\mathrm{E}\left[\epsilon_{\hat{\mathbfit{g}}}^2\right]&\geq\mathrm{E}\left[\mathcal{I}^{-1}(M,M,\zeta/\beta\tilde{\rho}_\mathrm{U})\right]\nonumber\\
&\geq\mathcal{I}^{-1}(M,M,\mathrm{E}[\zeta]/\beta\tilde{\rho}_\mathrm{U})\nonumber\\
& =\mathcal{I}^{-1}(M,M,M\tilde{\rho}_\mathrm{U})\quad (\mathrm{rad}^2). \label{eq:expUlCrb}
\end{align}

\begin{rem}
One should note that the CRB \eqref{eq:ul_crb_fixedG} relies on a truncated Taylor expansion. For the differentiation of the first derivatives on the Grassmannian, an additional structure (i.e., affine connection) has been introduced, which allows to connect tangent spaces at different points of the manifold. In \cite{1420804}, the sectional and Riemannian curvature terms that appear in the CRB because of the affine connection have been omitted for the sake of closed-form expressions. As we will illustrate in Section \ref{sec:monte_carlo}, the resulting CRB is legitimate only at a large enough SINR so that typical errors are small compared to the scale at which curvature becomes a dominant feature.
\end{rem}

\subsection{Downlink Subspace Estimation and CRB}\label{dlSsEst}
For the estimation of the DL subspace, we consider the conjugate transposed of the system model in \eqref{effSysModel},
\begin{align}
\tilde{\mathbfit{Y}}^H = \sqrt{\alpha}\underbrace{\left[\sqrt{\tau P/ M}\mathbfit{h}+\tilde{\mathbfit{w}}\right]}_{\tilde{\mathbfit{x}}}\mathbfit{g}^H + \tilde{\mathbfit{N}}^H, \label{effDlSysModel}
\end{align}
where $\mathbfit{h}$ is fixed but unknown, $\mathbfit{g}$ is a complex Gaussian random vector, and $\tilde{\mathbfit{x}}$ denotes an auxiliary variable which represents the pilot reverse modulated received signal at the repeater.
Analog to the previous section, we associate subspaces with points on the hypersphere, and exploit the independence of ML estimator for $[\mathbfit{h}]$ with respect to the norm of $\mathbfit{h}$.
Consequently, the ML estimate for the DL subspace, given the observation $\tilde{\mathbfit{Y}}$, is obtained by
\begin{align}
\hat{\mathbfit{h}}^\mathrm{ML} &\sim \mathrm{arg}\;\underset{\mathbfit{h}\in\mathbb{S}^{M-1}}{\mathrm{max}}\; p(\tilde{\mathbfit{Y}}|\mathbfit{h}),
\end{align}
where $p(\tilde{\mathbfit{Y}}|\mathbfit{h})$ denotes the conditional pdf of $\tilde{\mathbfit{Y}}$ given $\mathbfit{h}$.
Due to the existence of the term $\tilde{\mathbfit{w}}\mathbfit{g}^H$ in \eqref{effDlSysModel}, which follows a (degenerate) complex double Gaussian distribution \cite{6086771}, no closed-form expression for $p(\tilde{\mathbfit{Y}}|\mathbfit{h})$ is available. However, by introducing the auxiliary variable $\tilde{\mathbfit{x}}$, \eqref{effDlSysModel} forms the Markov chain
\begin{align}
\mathbfit{h} \overset{p(\tilde{\mathbfit{x}}|\mathbfit{h})}{\longrightarrow} \tilde{\mathbfit{x}} \overset{p(\tilde{\mathbfit{Y}}|\tilde{\mathbfit{x}})}{\longrightarrow} \tilde{\mathbfit{Y}}, \label{markovChain}
\end{align}
whose joint pdf is 
\begin{align}
p(\mathbfit{h},\tilde{\mathbfit{x}},\tilde{\mathbfit{Y}})=p(\mathbfit{h})p(\tilde{\mathbfit{x}}|\mathbfit{h})p(\tilde{\mathbfit{Y}}|\tilde{\mathbfit{x}}).
\end{align}
Consequently, the conditional pdf $p(\tilde{\mathbfit{Y}}|\mathbfit{h})$ for a fixed $\tilde{\mathbfit{Y}}$ can be written as the marginal distribution
\begin{align}
p(\tilde{\mathbfit{Y}}|\mathbfit{h})&=\int p(\tilde{\mathbfit{Y}},\tilde{\mathbfit{x}}|\mathbfit{h})  \, d\tilde{\mathbfit{x}}\nonumber\\
&\propto \int  p(\mathbfit{h},\tilde{\mathbfit{x}},\tilde{\mathbfit{Y}}) \,d\tilde{\mathbfit{x}}\nonumber\\
&\propto \int  p(\tilde{\mathbfit{x}}|\mathbfit{h})p(\tilde{\mathbfit{Y}}|\tilde{\mathbfit{x}}) \,d\tilde{\mathbfit{x}},
\end{align}
where we exploited the fact that $\mathbfit{h}$ and $\tilde{\mathbfit{x}}$ are isotropically distributed, and the irrelevance of the fixed probability $p(\tilde{\mathbfit{Y}})$ with respect to our inference problem.

The ML estimate for the subspace of $\mathbfit{h}$ is thus obtained by
\begin{align}
\hat{\mathbfit{h}}^\mathrm{ML} &\sim \mathrm{arg}\;\underset{\mathbfit{h}\in\mathbb{S}^{M-1}}{\mathrm{max}}\; \int p(\tilde{\mathbfit{x}}|\mathbfit{h})p(\tilde{\mathbfit{Y}}|\tilde{\mathbfit{x}}) \, d\tilde{\mathbfit{x}},\label{eq:dl_argmax}
\end{align}
where the likelihood function $p(\tilde{\mathbfit{x}}|\mathbfit{h})$ is
\begin{align}
p(\tilde{\mathbfit{x}}|\mathbfit{h}) = \left(\pi^M \mathrm{det} \mathbfit{R}_{\tilde{\mathbfit{x}}|\mathbfit{h}} \right)^{-1}  \mathrm{exp}\left\{- \tilde{\mathbfit{x}}^H\mathbfit{R}_{\tilde{\mathbfit{x}}|\mathbfit{h}}^{-1} \tilde{\mathbfit{x}}   \right\}, \label{eq:condProbhx}
\end{align}
with 
\begin{align}
\mathbfit{R}_{\tilde{\mathbfit{x}}|\mathbfit{h}}=\mathrm{E}\left[\tilde{\mathbfit{x}}\tilde{\mathbfit{x}}^H|\mathbfit{h}\right]=\tau/M P \mathbfit{h} \mathbfit{h}^H  + \mathbfit{I}_M.
\end{align}
Because of $\mathbfit{h}\in\mathbb{S}^{M-1}$, the likelihood function $p(\tilde{\mathbfit{x}}|\mathbfit{h})$ is again (cf. Section \ref{UlSec}) proportional to the complex Bingham distribution with the parameter matrix $\tilde{\mathbfit{x}}  \tilde{\mathbfit{x}} ^H$, whose maximum is attained by choosing $\mathbfit{h} \sim \tilde{\mathbfit{x}}$.
Consequently, the function $p(\tilde{\mathbfit{x}}|\mathbfit{h})$ can be replaced by $\delta(\mathbfit{h} - \tilde{\mathbfit{x}})$ without changing the maximizing argument in \eqref{eq:dl_argmax}; that is,
\begin{align}
\hat{\mathbfit{h}}^\mathrm{ML} &\sim \mathrm{arg}\;\underset{\mathbfit{h}\in\mathbb{S}^{M-1}}{\mathrm{max}}\; \int \delta(\mathbfit{h} - \tilde{\mathbfit{x}}) p(\tilde{\mathbfit{Y}}|\tilde{\mathbfit{x}})  d\tilde{\mathbfit{x}}\nonumber\\
& \sim \mathrm{arg}\;\underset{\tilde{\mathbfit{x}}\in\mathbb{S}^{M-1}}{\mathrm{max}}\; p(\tilde{\mathbfit{Y}}|\tilde{\mathbfit{x}}),
\end{align}
where the last step exploits $\tilde{\mathbfit{x}} \sim \mathbfit{h}$.
Finally, we need to find the subspace that maximizes the likelihood function $p(\tilde{\mathbfit{Y}}|\tilde{\mathbfit{x}})$. Analog to Section \ref{UlSec}, $\tilde{\mathbfit{Y}}$ given $\tilde{\mathbfit{x}}$ is zero-mean and Gaussian, and its conditional pdf is given by
\begin{align}
p(\tilde{\mathbfit{Y}}|\tilde{\mathbfit{x}})=\left(\pi^M \mathrm{det} \mathbfit{R}_{\tilde{\mathbfit{Y}}|\tilde{\mathbfit{x}}} \right)^{-M}  \mathrm{etr}\left\{- \mathbfit{R}_{\tilde{\mathbfit{Y}}|\tilde{\mathbfit{x}}}^{-1} \tilde{\mathbfit{Y}}^H\tilde{\mathbfit{Y}}  \right\},
\end{align}
where $\mathbfit{R}_{\tilde{\mathbfit{Y}}|\tilde{\mathbfit{x}}}$ denotes the conditional covariance matrix for an arbitrary row of $\tilde{\mathbfit{Y}}$,
\begin{align}
\mathbfit{R}_{\tilde{\mathbfit{Y}}|\tilde{\mathbfit{x}}}= \alpha\beta \tilde{\mathbfit{x}} \tilde{\mathbfit{x}}^H  + \mathbfit{I}_M.
\end{align}
Since we can again assume $\tilde{\mathbfit{x}}\in\mathbb{S}^{M-1}$, the computation of the determinant and the inverse of $\mathbfit{R}_{\tilde{\mathbfit{Y}}|\tilde{\mathbfit{x}}}$ shows that $p(\tilde{\mathbfit{Y}}|\tilde{\mathbfit{x}})$ is proportional to the Bingham distribution with the parameter matrix $\tilde{\mathbfit{Y}}^H\tilde{\mathbfit{Y}}$; that is,
\begin{align}
p(\tilde{\mathbfit{Y}}|\tilde{\mathbfit{x}})\propto  \mathrm{exp}\left\{\tilde{\mathbfit{x}}^H \tilde{\mathbfit{Y}}^H\tilde{\mathbfit{Y}} \tilde{\mathbfit{x}} \right\}.
\end{align}
The maximizing argument $\tilde{\mathbfit{x}} \sim \hat{\mathbfit{h}}^\mathrm{ML}$ is given by the dominant eigenvector $ \mathbfit{v}_\mathrm{max}(\tilde{\mathbfit{Y}}^H\tilde{\mathbfit{Y}})$ of $\tilde{\mathbfit{Y}}^H\tilde{\mathbfit{Y}}$; that is,
\begin{align}
\hat{\mathbfit{h}}^\mathrm{ML} \sim \mathbfit{v}_\mathrm{max}(\tilde{\mathbfit{Y}}^H\tilde{\mathbfit{Y}}). \label{DL_ML_EST}
\end{align}

The CRB for the DL subspace estimator consists of two components. First note that the (transformed) DL perturbations $\tilde{\mathbfit{w}}$ constitute a residual impairment because the ML subspace estimate $[\hat{\mathbfit{h}}^\mathrm{ML}]$ equals $[\tilde{\mathbfit{x}}]$. The MSE caused by this residual perturbation is lower bound by the CRB $\bar{\epsilon}_{\mathbfit{h},\mathrm{D}}^2$ of the estimator for $[\mathbfit{h}]$ given the single (DL) observation $\tilde{\mathbfit{x}}$. Again, we can apply the inverse FIM \eqref{eq:inverseFIM} with $T=1$, and the (average) effective SINR $\bar{\gamma}_{\mathbfit{h},\mathrm{D}} =   \mathrm{E}\left[\left\|{\mathbfit{h}}\right\|^2\right]/\beta\tilde{\rho}_\mathrm{D}=M\tilde{\rho}_\mathrm{D}$,
yielding the first CRB component
\begin{align}
\bar{\epsilon}_{\mathbfit{h},\mathrm{D}}^2 \geq \mathcal{I}^{-1}(M,1,M\tilde{\rho}_\mathrm{D}) \quad (\mathrm{rad}^2). \label{dlCrbInfRhoU}
\end{align}
The second subspace error component is due to the UL perturbation, which limits the estimation accuracy when estimating $\mathbfit{h} \sim \tilde{\mathbfit{x}}$ from the UL observation $\tilde{\mathbfit{Y}}$. Employing the inverse FIM \eqref{eq:inverseFIM} with $T=M$, and the (average) effective SINR $M\tilde{\rho}_\mathrm{U}$ (cf. Section \ref{UlSec}), we have the second CRB component
\begin{align}
\bar{\epsilon}_{\mathbfit{h},\mathrm{U}}^2 \geq \mathcal{I}^{-1}(M,M,M\tilde{\rho}_\mathrm{U}) \quad (\mathrm{rad}^2). \label{dlCrbInfRhoD}
\end{align}
Due to the independence of the UL \& DL perturbations, the resulting total CRB is given by the sum of the individual error variances; that is,
\begin{align}
\bar{\epsilon}_{\mathbfit{h}}^2 \geq \mathcal{I}^{-1}(M,1,M\tilde{\rho}_\mathrm{D}) +  \mathcal{I}^{-1}(M,M,M\tilde{\rho}_\mathrm{U}). \label{dlCrb}
\end{align}

\begin{rem}
A comparison of \eqref{eq:expUlCrb} with \eqref{dlCrb} shows that the DL CRB is the sum of the CRB for estimating $\mathbfit{h}$ from $\tilde{\mathbfit{x}}$, and the CRB for estimating $\tilde{\mathbfit{x}}$ from $\tilde{\mathbfit{Y}}$, which is identical to the CRB \eqref{eq:expUlCrb} of the UL subspace estimator. The DL subspace estimator suffers from the DL and UL perturbations, and is for finite $\tilde{\rho}_\mathrm{D}$ always inferior to the UL estimator, whose performance is only degraded by the UL perturbations. 
\end{rem}

% ##########################################################################

\subsection{Reduction of the Computational Complexity}\label{compl_reduc}
For the joint estimation of the UL \& DL subspaces, we need to compute the dominant eigenvectors $\mathbfit{v}_\mathrm{max}(\tilde{\mathbfit{Y}}\tilde{\mathbfit{Y}}^H)$ and $\mathbfit{v}_\mathrm{max}(\tilde{\mathbfit{Y}}^H\tilde{\mathbfit{Y}})$.
Both can be simultaneously obtained from the singular value decomposition (SVD)
\begin{align}
\tilde{\mathbfit{Y}} :=\mathbfit{U}\boldsymbol\Sigma \mathbfit{V}^H.\label{eq:svd}
\end{align}
The columns of $\mathbfit{U}=\left[\mathbfit{u}_1,\ldots,\mathbfit{u}_M\right]$ and $\mathbfit{V}=\left[\mathbfit{v}_1,\ldots,\mathbfit{v}_M\right]$ denote the left- and right-singular vectors of $\tilde{\mathbfit{Y}}$, and $\boldsymbol\Sigma$ is the diagonal matrix (sorted in descending order) of singular values $\left\{\sigma_m\right\}_{m=1}^M$.
The ML estimates for UL \& DL subspaces are given by left- and right-singular vectors that correspond to the largest singular value $\sigma_1$; that is,
\begin{align}
\hat{\mathbfit{g}}^\mathrm{ML}\sim \mathbfit{u}_1,\label{eq:ul_svd_est}\\
\hat{\mathbfit{h}}^\mathrm{ML}\sim \mathbfit{v}_1.\label{eq:dl_svd_est}
\end{align}

\subsubsection{SVD via Power Iterations}
The computational (time) complexity of the SVD algorithm for a square $M\times M$ matrix $\tilde{\mathbfit{Y}}$ is generally $\mathcal{O}(M^3)$; that is, it is not well-suited for high-dimensional matrices. However, the dominant left- and right-singular vectors of $\tilde{\mathbfit{Y}}$ that correspond to the largest singular value, can be easily approximated by using the power iteration method \cite[Section 7.3.1]{svd}. For example, the UL subspace estimate $\hat{\mathbfit{g}}^\mathrm{ML}$ is iteratively approximated by the recurrence relation
\begin{align}
\hat{\mathbfit{g}}_{i}={\tilde{\mathbfit{Y}}\tilde{\mathbfit{Y}}^H\hat{\mathbfit{g}}_{i-1}}/{\|\tilde{\mathbfit{Y}}\tilde{\mathbfit{Y}}^H\hat{\mathbfit{g}}_{i-1}\|}.\label{eq:pwrIt}
\end{align}
If we assume that the matrix $\tilde{\mathbfit{Y}}\tilde{\mathbfit{Y}}^H$ has an eigenvalue that is strictly greater in magnitude than its other eigenvalues, and the starting vector $\hat{\mathbfit{g}}_{0}$ has a nonzero dot product with the dominant eigenvector of $\tilde{\mathbfit{Y}}\tilde{\mathbfit{Y}}^H$, then a subsequence $(\hat{\mathbfit{g}}_{i})$ converges to the true dominant eigenvector.
As a simple stopping criterion, we propose using the distance $d_\mathrm{ss}(\hat{\mathbfit{g}}_{i},\hat{\mathbfit{g}}_{i-1})$ between two consecutively computed subspaces, which is required to be smaller than a predefined threshold $\delta$. For a given $\hat{\mathbfit{g}}_{i}$, the corresponding left-singular vector $\hat{\mathbfit{h}}_i$ and the singular value $\hat{\sigma}_1$ are calculated by 
\begin{align}
\hat{\mathbfit{h}}_i &= {\tilde{\mathbfit{Y}}^H\hat{\mathbfit{g}}_{i}}/{\|\tilde{\mathbfit{Y}}^H\hat{\mathbfit{g}}_{i}\|}\label{eq:pwrIt2},\\
\hat{\sigma}_1 &= |\hat{\mathbfit{g}}_i^H \tilde{\mathbfit{Y}}\hat{\mathbfit{h}}_{i}|. \label{eq:pwrIt3}
\end{align}
If the number of required power iterations is smaller than $\frac{M^2}{M+1}+1$, then the joint complexity (in terms of dot products) of the computations \eqref{eq:pwrIt}, \eqref{eq:pwrIt2} can be reduced by splitting the matrix-vector product in \eqref{eq:pwrIt} (which requires a non-recurring matrix-matrix multiplication) into two successive steps,
\begin{align}
\hat{\mathbfit{h}}_{i}&={\tilde{\mathbfit{Y}}^H\hat{\mathbfit{g}}_{i-1}}/{\|\tilde{\mathbfit{Y}}^H\hat{\mathbfit{g}}_{i-1}\|},\\
\hat{\mathbfit{g}}_{i}&={\tilde{\mathbfit{Y}}\hat{\mathbfit{h}}_{i}}/{\|\tilde{\mathbfit{Y}}\hat{\mathbfit{h}}_{i}\|}.
\end{align}
One should note that the number of required iterations strongly depends on the initial value $\hat{\mathbfit{g}}_{0}$. Therefore, we propose to use the column of $\tilde{\mathbfit{Y}}=[\tilde{\mathbfit{y}}_1,\ldots,\tilde{\mathbfit{y}}_M]$ with the largest Euclidean norm because it has the smallest (relative) noise perturbation with a high probability. The overall procedure is summarized in Algorithm \ref{pwrIt}, where we extended the stopping criterion by the DL subspace increment.

\begin{algorithm}[t]
\caption{Rank-1 SVD based on Power Iterations}\label{pwrIt}
\begin{algorithmic}[1]
\Procedure{PowerIteration}{$\tilde{\mathbfit{Y}},\delta$}
\State $m^* \gets \mathrm{arg}\,\underset{m\in[1,M]}{\mathrm{max}}\; \left\|\tilde{\mathbfit{y}}_m\right\|^2$ 
\State $\hat{\mathbfit{g}}_{0} \gets \slfrac{\tilde{\mathbfit{y}}_{m^*}}{\left\|\tilde{\mathbfit{y}}_{m^*}\right\|}$ 
\State $\hat{\mathbfit{h}}_{0} \gets M^{-\frac{1}{2}}\boldsymbol 1_M$ 
\State $i\gets 0$
\Repeat
\State $i\gets i+1$
\State $\hat{\mathbfit{h}}_i \gets {\tilde{\mathbfit{Y}}^H \hat{\mathbfit{g}}_{i-1}}/{\|\tilde{\mathbfit{Y}}^H \hat{\mathbfit{g}}_{i-1}\|}$
\State $\hat{\mathbfit{g}}_i \gets {\tilde{\mathbfit{Y}}\hat{\mathbfit{h}}_{i}}/{\|\tilde{\mathbfit{Y}}\hat{\mathbfit{h}}_{i}\|}$
\Until{$(\mathrm{max}\{d_\mathrm{ss}(\hat{\mathbfit{g}}_{i-1},\hat{\mathbfit{g}}_i),d_\mathrm{ss}(\hat{\mathbfit{h}}_{i-1},\hat{\mathbfit{h}}_i)\} \leq \delta)$} 
\State $\hat{\sigma}_1 \gets |\hat{\mathbfit{g}}_i^H \tilde{\mathbfit{Y}}\hat{\mathbfit{h}}_{i}|$
\State \textbf{return} $\hat{\mathbfit{h}}_i$, $\hat{\mathbfit{g}}_i,\hat{\sigma}_1$
\EndProcedure
\end{algorithmic}
\end{algorithm}

\subsubsection{Shifting/Omitting the Pilot Reverse Modulation Step}
For the UL subspace estimation with $\tau/M=1$, the pilot reverse modulation step in \eqref{effSysModel} can be omitted because the individual left-singular vectors $\mathbfit{U}=\left[\mathbfit{u}_1,\ldots,\mathbfit{u}_M\right]$ of $\tilde{\mathbfit{Y}}:=\mathbfit{U}\boldsymbol\Sigma \mathbfit{V}^H$ and
$\mathbfit{Y}:=\mathbfit{U}\boldsymbol\Sigma \tilde{\mathbfit{V}}^H=\mathbfit{U}\boldsymbol\Sigma \mathbfit{V}^H\boldsymbol{\Phi}^H$
span the same subspaces. 

Similarly for the DL subspace estimation with $\tau/M=1$, the order of the pilot reverse modulation step and the SVD-based DL subspace estimation can be swapped; that is, 
\begin{align}
\hat{\mathbfit{h}} \sim \boldsymbol\Phi^H \tilde{\mathbfit{v}}_1.
\end{align}
where $\tilde{\mathbfit{v}}_1$ denotes the right-singular vector of $\mathbfit{Y}:=\mathbfit{U}\boldsymbol\Sigma \left[\tilde{\mathbfit{v}}_1,\ldots,\tilde{\mathbfit{v}}_M\right]^H$, with $\mathbfit{Y}$ given in \eqref{eq:ul_rcvd_sgnl}. Note that we have $\tilde{\mathbfit{v}}_1=\boldsymbol\Phi\mathbfit{v}_1$ with $\mathbfit{v}_1$ given in \eqref{eq:svd}. Consequently, the matrix-matrix multiplication in \eqref{effSysModel} can be reduced to a matrix-vector multiplication, which incurs no performance loss for the case $\tau/M=1$.

% ##########################################################################

\section{Uplink Channel Gain Estimation}\label{sec:ulChanNormEst}
For the power control of data streams, knowledge of the UL \& DL channel gains is required. Ideally, these gains are estimated separately, but an inspection of the feedback signal in \eqref{effSysModel} reveals that without any knowledge of the DL perturbation variance, the AA cannot derive an estimate for the DL channel vector norm. Since we assume fully analog repeaters, there exists no simple way to estimate and feedback DL perturbation variance to the AA. Consequently, we focus on the estimation of the UL channel vector norm, and exploit the fact that on average the DL frequency response has approximately the same power as the UL frequency response, so that it is reasonable to assume them to be equal.

\subsection{Maximum Likelihood Estimation}
Again, we split the UL channel vector into two independent random variables (cf. \eqref{eq:ulVecSplit}). The reverse-modulated UL signal \eqref{effSysModel} becomes
\begin{align}
\tilde{\mathbfit{Y}} =\sqrt{\alpha\zeta} \boldsymbol\phi \tilde{\mathbfit{x}}^H + \tilde{\mathbfit{N}},
\end{align}
where $\tilde{\mathbfit{x}}=\boldsymbol{\Phi}^H\mathbfit{x}$.
In the following, we are interested in the ML estimate for $\zeta$, in presence of the unknown
\begin{itemize}
	\item isotropically distributed complex unit vector $\boldsymbol\phi\in\mathbb{S}^{M-1}$,
	\item complex Gaussian signal $\sqrt{\alpha}\tilde{\mathbfit{x}}\in\mathbb{C}^M$ with i.i.d. entries of variance $\tilde{Q}=\alpha(\tau/M\beta P+1)$,
	\item complex Gaussian perturbation $\tilde{\mathbfit{N}}\in\mathbb{C}^{M\times M}$ with i.i.d. entries of unit variance.
\end{itemize}
We assume that the AA knows the (effective) transmit power $\tilde{Q}$ of the repeater, and the variance of the UL perturbations.
The ML estimate for $\zeta$ is defined as
\begin{align}
\hat{\zeta}^\mathrm{ML} &= \mathrm{arg}\;\underset{\zeta\in\mathbb{R}_+}{\mathrm{max}}\; p(\tilde{\mathbfit{Y}}|\zeta).\label{eq:argmaxChangain}
\end{align}
The likelihood function $p(\tilde{\mathbfit{Y}}|\zeta)$ can be written as the marginal pdf,
\begin{align}
p(\tilde{\mathbfit{Y}}|\zeta) = \int{p(\tilde{\mathbfit{Y}}|\boldsymbol\phi,\zeta)p(\boldsymbol\phi)} d\boldsymbol\phi, \label{eq:idInt}
\end{align}
where $p(\tilde{\mathbfit{Y}}|\boldsymbol\phi,\zeta)$ and $p(\boldsymbol\phi)$ are given by  \eqref{eq:pdfUL} and \eqref{eq:pdfPhi}, repsectively.
With $q(\zeta)={\zeta\tilde{Q}}/(1+\zeta\tilde{Q})$, the pdf in \eqref{eq:idInt} becomes
\begin{align}
p(\tilde{\mathbfit{Y}}|\zeta) \propto &\frac{1}{(1+\zeta\tilde{Q})^M}\int{\mathrm{etr}\left\{q(\zeta)\boldsymbol\phi^H \tilde{\mathbfit{Y}}\tilde{\mathbfit{Y}}^H  \boldsymbol\phi \right\}   p(\boldsymbol\phi)} d\boldsymbol\phi\nonumber\\
=&\frac{1}{(1+\zeta\tilde{Q})^M}\int{\mathrm{etr}\left\{q(\zeta)\boldsymbol\phi^H \mathbfit{U}\boldsymbol{\Sigma}^2\mathbfit{U}^H  \boldsymbol\phi \right\}   p(\boldsymbol\phi)} d\boldsymbol\phi\nonumber\\
=&\frac{1}{(1+\zeta\tilde{Q})^M}\int{\mathrm{etr}\left\{q(\zeta)\sum_{m=1}^M\left|\phi_m\right|^2\sigma_m^2\right\}p(\boldsymbol\phi)} d\boldsymbol\phi \label{eq:idInt2}
\end{align}
where we use the eigendecomposition $\tilde{\mathbfit{Y}}\tilde{\mathbfit{Y}}^H=\mathbfit{U}\boldsymbol{\Sigma}^2\mathbfit{U}^H$ (cf. \eqref{eq:svd}), and $\left\{\sigma_m^2\right\}_{m=1}^M$ denote the diagonal elements of $\boldsymbol{\Sigma}^2$. In the above, we change the integration variable from $\boldsymbol\phi$ to $\mathbfit{U}^H \boldsymbol\phi$, and use the fact that $\mathbfit{U}^H \boldsymbol\phi$ has the same probability density as $\boldsymbol\phi$.
Since we have almost surely $\sigma_1^2>\sigma_2^2>\ldots >\sigma_M^2>0$, a closed-form expression \cite[Section B.1]{Marzetta99capacityof} for \eqref{eq:idInt2} is given by
\begin{align}
p(\tilde{\mathbfit{Y}}|\zeta) \propto &\frac{1}{(1+\zeta\tilde{Q})^M} \sum_{m=1}^M \frac{e^{q(\zeta)\sigma_m^2}-e^{q(\zeta)\sigma_1^2}}{\prod_{n\neq m}q(\zeta)(\sigma_m^2-\sigma_n^2)}.\label{eq:closedfromInt}
\end{align}
Apparently, there is no closed-form expression for the argument $\hat{\zeta}^\mathrm{ML}$ that maximizes \eqref{eq:closedfromInt}.
By the problem structure, however, \eqref{eq:closedfromInt} has a unique maximum, which can be found e.g., by the Golden section method \cite[Appendix C.3]{bertsekas1995nonlinear}.

\begin{figure*}[t]
\centering
\includegraphics[width=9.0cm]{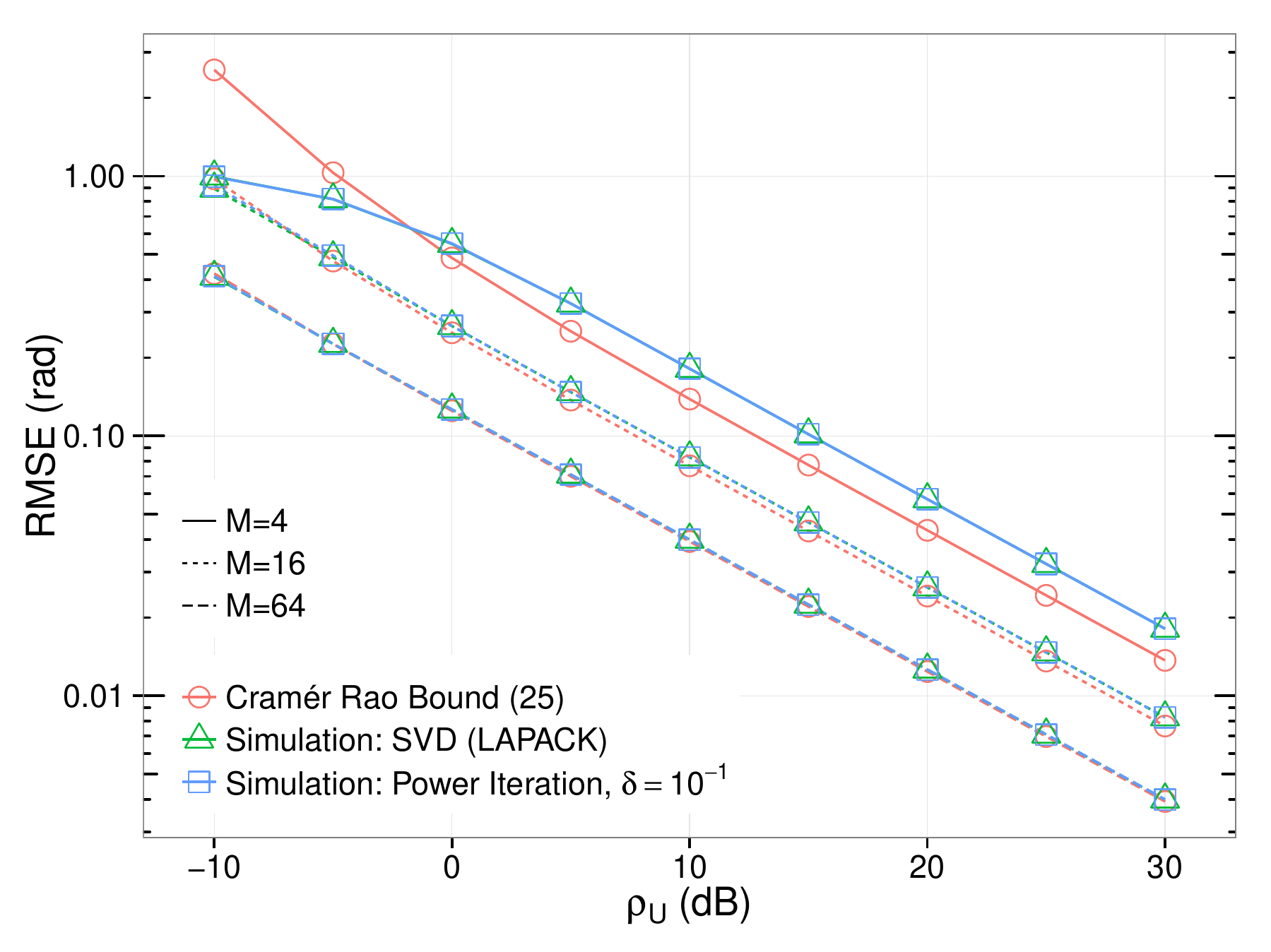}
\includegraphics[width=9.0cm]{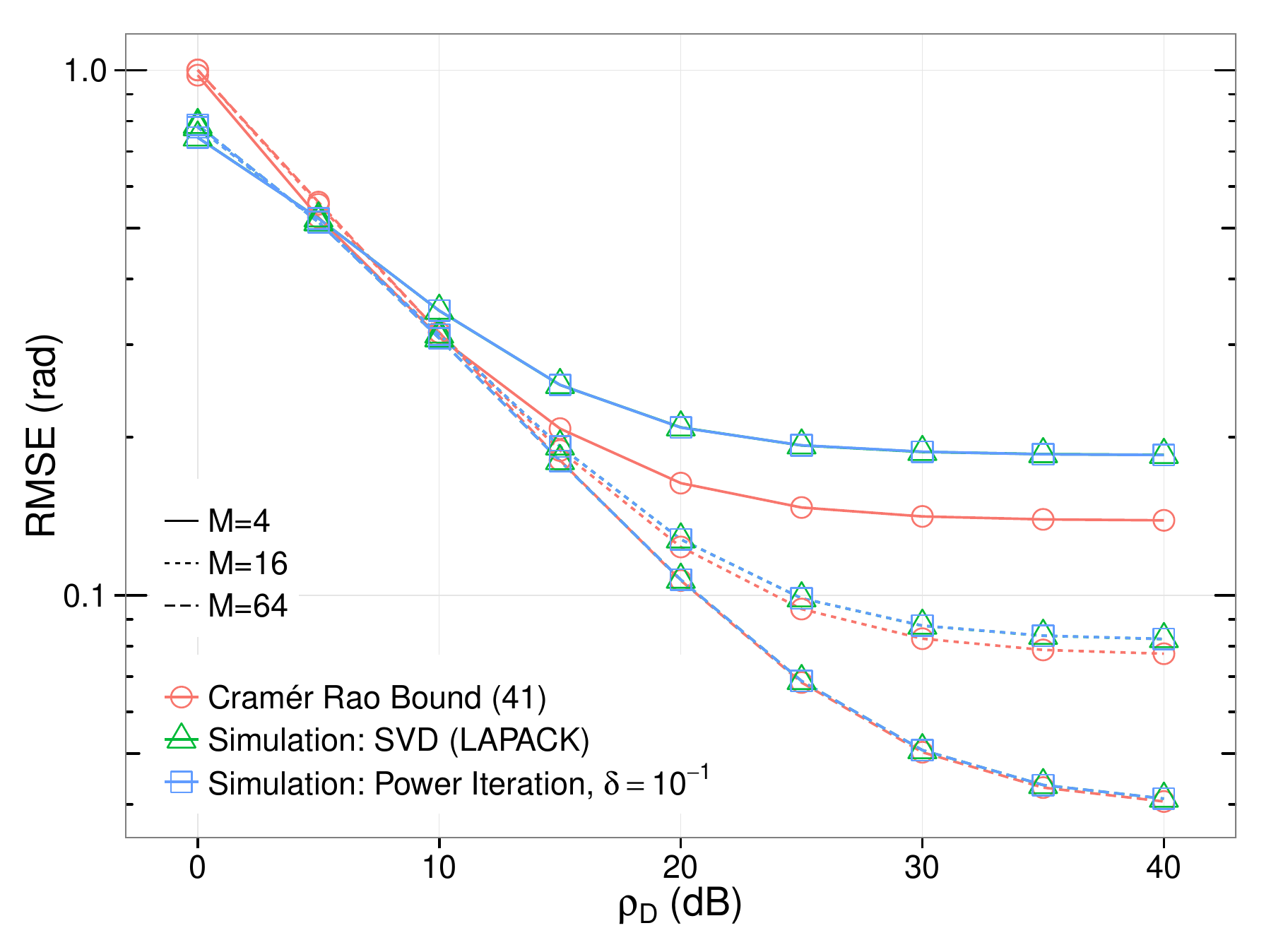}
\caption{Comparison between the \{SVD, Power Iteration\}-based subspace estimator's RMSE and the CRB for $M=\{4,16,64\}, \tau=M$; a) in the left figure as a function of the UL SINR $\rho_\mathrm{U}$, b) in the right figure as a function of the DL SINR $\rho_\mathrm{D}$ with fixed $\rho_\mathrm{U}=10$dB. Observe that the accuracy of the SVD-based estimator asymptotically approaches the CRB as $M(=\tau)\rightarrow\infty$; that is, the estimator is unbiased and asymptotically efficient. Moreover, one should note that the CRB fails for low SINRs (i.e., $\rho_\mathrm{U}<0$dB, $\rho_\mathrm{D}<10\mathrm{lg}(M)$dB) because it relies on truncated Taylor expansions.}
\label{fig:rmse}
\end{figure*}

\subsection{SCM-based Estimation}
The structure of the covariance matrix $\mathbfit{R}_{\tilde{\mathbfit{Y}}|\mathbfit{g}}$ in \eqref{eq:ulCovMat} reveals an alternative approach for the UL channel norm estimation. Its eigenvalues are given by $\left\{\zeta \tilde{Q}+1,1,\ldots,1\right\}$.
So, by computing the sample covariance matrix (SCM) 
\begin{align}
\hat{\mathbfit{R}}_{\tilde{\mathbfit{Y}}|\mathbfit{g}}=M^{-1}\tilde{\mathbfit{Y}}\tilde{\mathbfit{Y}}^H,
\end{align}
an estimate for $\zeta$ can be derived from its largest eigenvalue $\lambda_\mathrm{max}(\hat{\mathbfit{R}}_{\tilde{\mathbfit{Y}}|\mathbfit{g}})$; that is,
\begin{align}
\hat{\zeta} = \frac{\lambda_\mathrm{max}(\hat{\mathbfit{R}}_{\tilde{\mathbfit{Y}}|\mathbfit{g}})-1}{\tilde{Q}}=\frac{{\sigma}_1^2-M}{\tilde{Q}M},
\end{align}
where ${\sigma}_1$ denotes the largest singular value in \eqref{eq:svd}.
The accuracy of this estimate strongly depends on the accuracy of the SCM, which is only asymptotically efficient (i.e.,  for large values of $M$). For low sample support, the SCM is biased and not efficient with respect to the natural covariance metric \cite[Section III.C]{1420804}.

\begin{rem}
The same estimator is obtained by 
\begin{align}
\hat{\zeta} &= \mathrm{arg}\;\underset{\zeta\in\mathbb{R}_+}{\mathrm{max}}\; p(\tilde{\mathbfit{Y}}|\boldsymbol\phi=\hat{\mathbfit{g}}^\mathrm{ML},\zeta)\\
&=\mathrm{arg}\;\underset{\zeta\in\mathbb{R}_+}{\mathrm{max}}\; \frac{1}{(1+\zeta\tilde{Q})^M} e^{q(\zeta){\sigma}_1^2}\\
&=\frac{{\sigma}_1^2-M}{\tilde{Q}M}, \nonumber
\end{align}
where $p(\tilde{\mathbfit{Y}}|\boldsymbol\phi,\zeta)$ is given in \eqref{eq:pdfUL}, and $\hat{\mathbfit{g}}^\mathrm{ML}$ is the ML estimate for the UL subspace given in \eqref{eq:ul_ML_est}.
\end{rem}

% ##########################################################################

\begin{figure}[t]
\centering
\includegraphics[width=9.0cm]{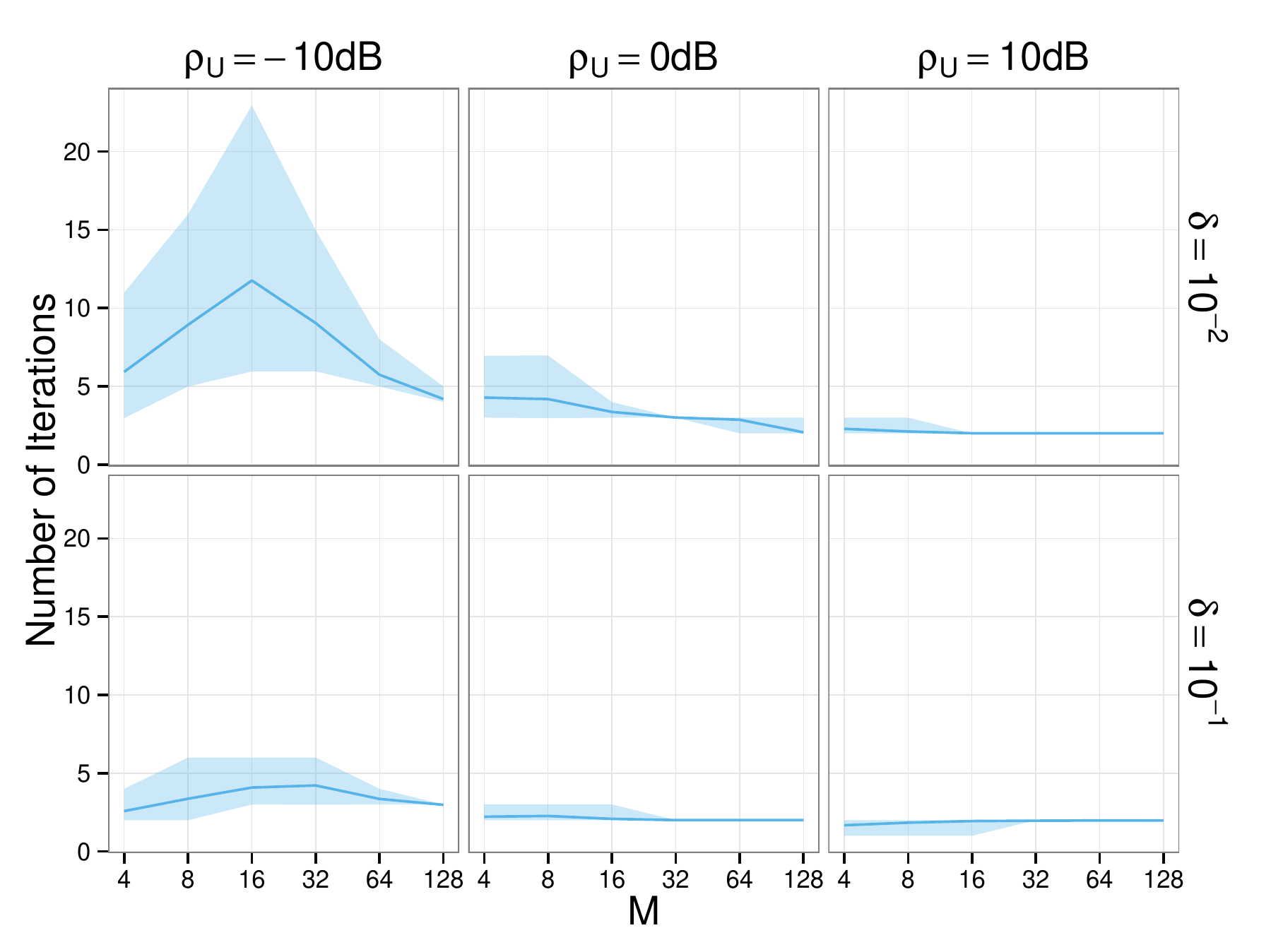}
\caption{Number of power iterations as a function of the number of array antennas $M$, for different UL SINRs $\rho_\mathrm{U}$ and accuracy bounds $\delta$. The solid line depicts the average number of iterations. The shade area illustrates the 5th- and 95th-percentiles of the iteration number's CDF. For low UL SINRs, the number required iterations grows because the distance between largest and second largest eigenvalues decreases.}
\label{fig:num_iterations}
\end{figure}

\begin{figure*}[t]
\centering
\includegraphics[width=9.0cm]{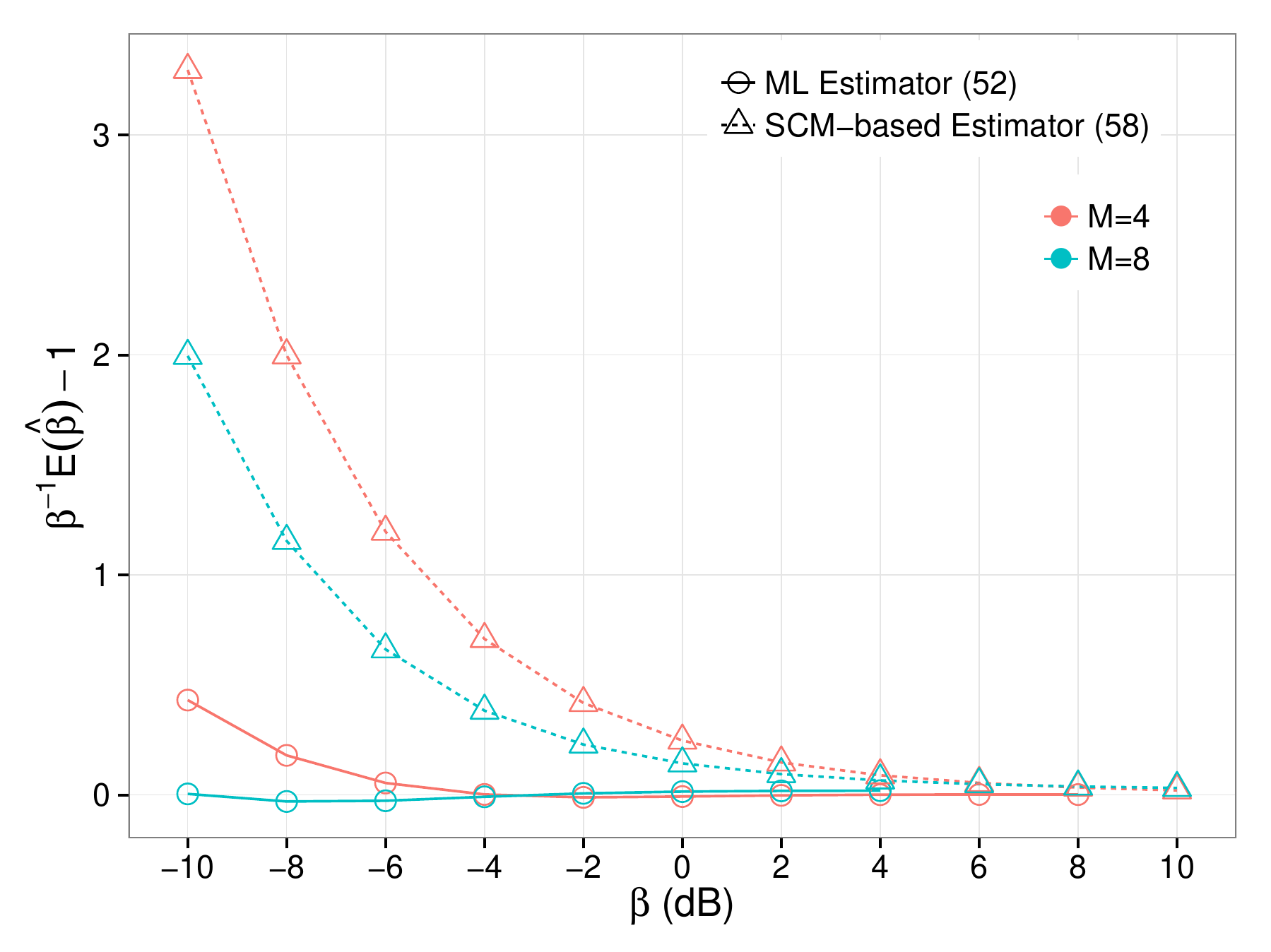}
\includegraphics[width=9.0cm]{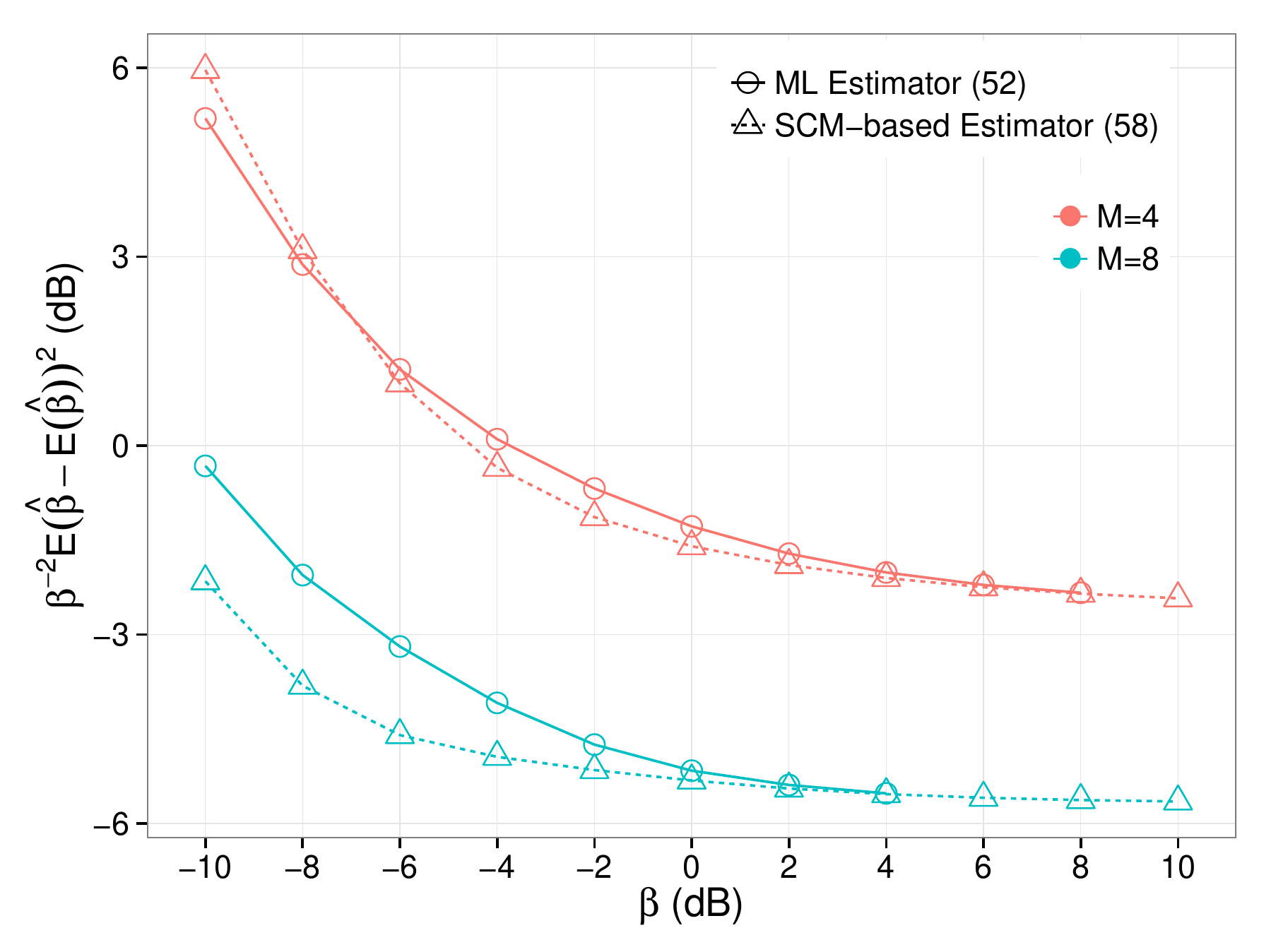}
\caption{Relative bias (left figure) and relative error variance (right figure) of the ML and SCM-based UL channel gain estimators, as a function of the channel gain $\beta$ for $Q=1$. The left figure shows that the ML estimator (solid lines) is unbiased for a sufficiently high number of antennas/observations $M\geq 8$. The SCM-based estimator (dashed lines) is biased for low $\beta=\rho_\mathrm{U}$ because the SCM is biased itself for low sample support. In terms of relative error variance, both estimators exhibit similar performance, which improves for an increasing number of antennas/observations $M$.}
\label{fig:beta}
\end{figure*}

\section{Numerical Illustration} \label{sec:monte_carlo}
Our analytical results in Section \ref{sec:ss_est} are corroborated by Monte Carlo simulations, which compare the subspace estimation accuracies of SVD-based estimator and the power iteration method against the CRBs.
In our simulations, we set $\tau=M$ and assume genie power control at the repeater side. Therefore, the repeater exploits perfect knowledge of the UL perturbation variance $\beta_\mathrm{U}$ in order to adjust its scaling factor 
\begin{align}
\alpha= \frac{\rho_\mathrm{U}}{\beta(\beta P+1)},
\end{align}
such that a predefined UL SINR $\rho_\mathrm{U}$ is achieved.

Given a parameter set $\{M,\rho_\mathrm{D},\rho_\mathrm{U}\}$, the CRBs are computed according to \eqref{eq:expUlCrb} and \eqref{dlCrb}, and 1000 Monte Carlo trials are performed. For each trial, the unitary pilot matrix $\boldsymbol{\Phi}$ is chosen randomly from the uniform distribution on the Stiefel manifold $V_M(\mathbb{C}^\tau)$, whereas the UL \& DL channel vectors and perturbations are generated according to Section \ref{sec:prob_form}.

The left side of Figure \ref{fig:rmse} shows the achieved UL root mean-square error (RMSE), which is independent of $\rho_\mathrm{D}$ because of $\tau=M$, and right side shows the DL RMSEs (for a fixed $\rho_\mathrm{U}=10$dB) as a function of the UL and DL SINRs, respectively. 
In both figures, one can observe a decreasing subspace estimation error as the number of array antennas $M$ grows. The UL estimation performance improves because the estimator benefits from an increased number of observations (i.e., feedback symbols), which grows with $M$. In the right figure, the DL estimation accuracy improves because it was offset by the limited SINR in the UL, which improves with a larger number of receive antennas $M$, each providing an independent observation of the same feedback signal. The benefit of this multi-antenna effect has already been noted in \cite{1608543}.
In addition, one can observe that the SVD-based estimator (green curve) is unbiased and asymptotically efficient; that is, its accuracy approaches asymptotically the CRB (red curve) as the number of observations $M(=\tau)$ becomes large (cmp. \cite{1420804}). On the other hand, the CRB fails as a lower bound for low SINRs because it does not incorporate the sectional and Riemannian curvature terms as argued in \cite[Section IV]{6418045}. Neglecting these terms can be legitimized for small errors only. For our system model, the CRB's range of validity is lower bounded by $\rho_\mathrm{U}>0$dB and $\rho_\mathrm{D}>10\mathrm{lg}(M)$dB. In addition, the blue curves in Figure \ref{fig:rmse} display the subspace RMSE for the power iteration based estimator, using the stopping criterion $\delta=10^{-1}$. Over typical ranges of the SINR parameters, the performance difference with respect to the SVD-based method is negligible.

In order to assess the computational complexity of the power iteration method, Figure \ref{fig:num_iterations} shows the average number of power iterations as a function of $M$, for various parameter combinations of $\rho_\mathrm{U}$ and $\delta$. In each subplot, we depict the shape the iteration number's cumulative distribution function (CDF) by a shaded area, which is lower and upper bounded by the 5th- and 95th-percentile of the CDF, respectively. The solid blue curve depicts the average number of iterations. We see that the number of required iterations decreases for increasing $M$ and $\rho_\mathrm{U}$; that is, the number of iterations can be upper bounded for all values of $M$ and $\rho_\mathrm{U}$, which yields a quadratic time complexity with respect to $M$. 
One should note that for low SINRs, the distance between the largest and second-largest eigenvalue of $\tilde{\mathbfit{Y}}\tilde{\mathbfit{Y}}^H$ decreases, which results in a slower convergence. For large $M$, we can observe a faster convergence of the power iterations. The UL estimation accuracy increases with growing $M$ because more UL observations are available (cmp. Figure \ref{fig:rmse}). A faster convergence of the left-singular vector in Algorithm \ref{pwrIt} will also result in a faster convergence of right-singular vector.

Finally, we analyze the relative bias and error variance for the UL channel gain estimators described in Section \ref{sec:ulChanNormEst}, assuming $Q=1$ which yields $\beta=\rho_\mathrm{U}$. Due to the exponential terms in \eqref{eq:closedfromInt}, the ML estimator is numerically intractable for large $M$ and $\beta$. Therefore, we omit all data points that exhibit numerical overflows.
The left side of Figure \ref{fig:beta} shows the relative bias of the estimators. For a sufficient number (i.e., $M\geq 8$) of UL observations, the ML estimator \eqref{eq:argmaxChangain} is unbiased for all simulated values of $\beta=\rho_\mathrm{U}$. This is quite remarkable because for $\rho_\mathrm{U}=-10$dB the signal component in the observation is 10 times weaker than the perturbations. In contrast, the SCM-based estimator exhibits a significant bias for small $\rho_\mathrm{U}$, which decreases as the number of available observations $M$ grows. Regarding the relative error variance as shown on the right side of Figure \ref{fig:beta}, both estimators exhibit a similar performance, which improves with growing $M$ because of the number of UL observations increases.

% ##########################################################################

\section{Summary and Discussion}\label{sec:sum_disc}
For the UL subspace estimation, we have shown that the subspace which maximizes the likelihood function is given by the dominant eigenvector of the sample covariance matrix (SCM) for the antenna array's observations.
For the DL subspace, no closed-form expression for the likelihood function exists due to the presence of a degenerate double Gaussian term. However, by exploiting the Markovian property of the signal model, we could show that the problem of finding the maximum likelihood estimate reduces to the maximization of the likelihood function for the received signal at the repeater. The solution is given by the dominant eigenvector of the sample covariance matrix for the conjugate transposed of the antenna arrays's observation matrix. Consequently, both estimates can be obtained simultaneously by computing the SVD of the pilot-reverse modulated receive signal. This result is all the more intuitive because we assumed spatially and temporally white perturbations, and also because the SVD is the optimal estimator for the noise-free case.

Based on the inverse Fisher information metric (FIM) from \cite{1420804}, we have formulated the CRBs for the unbiased UL and DL subspace estimators, and demonstrated the asymptotic efficiency of the SVD-based estimators by means of Monte Carlo simulations. Since the inverse FIM relies on truncated Taylor expansions, the formulated CRBs hold only for small perturbations; that is, their range of validity is lower bounded by $\rho_\mathrm{U}>0$dB and $\rho_\mathrm{D}>10\mathrm{lg}(M)$dB.

In addition, we have discussed several possibilities for the reduction of the processing complexity. For the case $\tau/M=1$, the order of the pilot reverse modulation and the SVD-based DL subspace estimator can be swapped without any performance penalties, which turns a matrix-matrix multiplication into a matrix-vector multiplication. For the UL subspace estimation with $\tau/M=1$, the pilot reverse modulation can be completely omitted. Moreover, the SVD computation, which has generally a cubic time complexity, can be replaced by the power iteration method. As shown by our simulations, the number of iterations can be kept constant with respect to the number of antennas $M$, which implies a quadratic time complexity. 

Finally, we formulated the ML estimator for the UL channel gain, which relies on the knowledge of the repeater's average transmit power and the UL perturbation variance. The resulting likelihood function, however, does not admit a closed-form solution for its maximum, so that we have to resort to a numerical line search method. As shown by the numerical experiments, this ML estimator is robust against strong UL perturbations (i.e., it is unbiased for a sufficiently large number of antennas/observations), but is numerically unstable for large $M$ and $\rho_\mathrm{U}$. Therefore, we propose a simple SCM-based estimator, which is naturally biased for a small sample support (i.e., small values of $M$), but exhibits a similar relative error variance as the ML estimator.

In the following, we briefly discuss some open research questions.

\subsubsection{ML Estimation for Nonwhite Perturbations}
In \cite{1420804}, the case of correlated noise with a known covariance matrix is discussed. In our UL model \eqref{effSysModel}, this corresponds to a perturbation matrix $\tilde{\mathbfit{N}}=[\tilde{\mathbfit{n}}_1,\ldots,\tilde{\mathbfit{n}}_\tau]$ with i.i.d. columns, each following $\mathcal{CN}(\boldsymbol{0}_M,\mathbfit{R}_{\tilde{\mathbfit{n}}})$, where $\mathbfit{R}_{\tilde{\mathbfit{n}}}=\mathrm{E}[\tilde{\mathbfit{n}}_t\tilde{\mathbfit{n}}_t^H],\forall t$ denotes the $M \times M$ covariance matrix of the transformed perturbations. Note that this covariance model admits only sensible interpretations for $\tau/M=1$, because then we have 
$\mathbfit{R}_{\tilde{\mathbfit{n}}}=\mathrm{E}[\mathbfit{n}_t\boldsymbol{\Phi}\boldsymbol{\Phi}^H\mathbfit{n}_t^H]=\mathrm{E}[\mathbfit{n}_t\mathbfit{n}_t^H],\forall t$; that is, we observe temporally independent (w.r.t. the index $t$) but spatially correlated perturbations $\mathbfit{N}=[{\mathbfit{n}}_1,\ldots,{\mathbfit{n}}_\tau]$ in \eqref{eq:ul_rcvd_sgnl}.
For nonwhite noise, the simple SVD-based estimator in \eqref{eq:ul_svd_est} is biased by the principal invariant subspace of $\mathbfit{R}_{\tilde{\mathbfit{n}}}$. Therefore, a whitened SVD-based approach is proposed
\begin{align}
\mathbfit{R}_{\tilde{\mathbfit{n}}}^{-1/2}\tilde{\mathbfit{Y}} :=\mathbfit{U}\boldsymbol\Sigma \mathbfit{V}^H,\label{eq:white_svd}
\end{align}
where the ML estimate for $\mathbfit{g}$ is given by $\mathbfit{g}^\mathrm{ML}\sim \mathbfit{R}_{\tilde{\mathbfit{n}}}^{1/2}\mathbfit{u}_1$.
One should note that this spatial covariance model for the UL perturbations implies ``temporally'' correlated UL perturbations (i.e., $\tilde{\mathbfit{N}}^H$ in \eqref{effDlSysModel} with correlated columns) for the DL subspace estimator, which violates our model assumption even for the asymptotic case of $\rho_\mathrm{D}\rightarrow\infty$. The joint pdf of the observation matrix $\tilde{\mathbfit{Y}}^H$ cannot be written as the product of the columns' pdfs. The derivation of the corresponding ML estimator as well as its CRB is subject to future work.

\subsubsection{Bayesian Subspace Estimation}\label{sec:bayesian}
A Bayesian subspace estimation approach is described in \cite{6006540}, which is applicable to our UL subspace estimation for the case $\tau/M=1$ (i.e., the elements of $\mathbfit{x}$ in \eqref{eq:ul_rcvd_sgnl} must be i.i.d.). The resulting estimator is called minimum mean square distance (MMSD) subspace estimator because the MSE in \eqref{eq:mse_def} is adopted as the cost function. A closed-form solution is derived for the Bingham distribution \cite{complBingham} as the pior for $\mathbfit{g}$, given by
\begin{align}
\pi(\mathbfit{g})\propto \mathrm{etr}\{\kappa \mathbfit{g}^H \boldsymbol\Pi \boldsymbol\Pi^H \mathbfit{g} \},
\end{align}
where $\boldsymbol\Pi\in G_{p}(\mathbb{C}^M)$ reflects the prior knowledge about the $p$-dimensional subspace where $\mathbfit{g}$ evolves, and $\kappa$ is a concentration parameter: the larger $\kappa$ the more concentrated around $\boldsymbol\Pi$ is the subspace of $\mathbfit{g}$. 
The MMSD estimate $\mathbfit{g}^\mathrm{MMSD}$ is given by the dominant eigenvector of the matrix $(\kappa\boldsymbol\Pi\boldsymbol\Pi^H + \mathbfit{Y}\mathbfit{Y}^H)$.
Obviously, the MMSD estimator collapses to the ML estimator for the case of an uniform prior distribution (i.e., when $\mathbfit{g}$ is isotropically distributed).
Unfortunately, a MMSD estimator for the DL subspace cannot be deduced from the results provided in \cite{6006540}, and is subject to future work.

% ##########################################################################
\section*{Acknowledgment}
This work has been conducted within Bell Labs' F-Cell project. The authors wish to thank project initiators T. Klein, T. Sizer, A. Pascht and O. Blume. They are also grateful for the helpful comments of V. Suryaprakash and H. Schlesinger.

\bibliographystyle{IEEEtran}
\bibliography{IEEEabrv,bibtex_refs}

\begin{thebibliography}{10}
\providecommand{\url}[1]{#1}
\csname url@samestyle\endcsname
\providecommand{\newblock}{\relax}
\providecommand{\bibinfo}[2]{#2}
\providecommand{\BIBentrySTDinterwordspacing}{\spaceskip=0pt\relax}
\providecommand{\BIBentryALTinterwordstretchfactor}{4}
\providecommand{\BIBentryALTinterwordspacing}{\spaceskip=\fontdimen2\font plus
\BIBentryALTinterwordstretchfactor\fontdimen3\font minus
  \fontdimen4\font\relax}
\providecommand{\BIBforeignlanguage}[2]{{%
\expandafter\ifx\csname l@#1\endcsname\relax
\typeout{** WARNING: IEEEtran.bst: No hyphenation pattern has been}%
\typeout{** loaded for the language `#1'. Using the pattern for}%
\typeout{** the default language instead.}%
\else
\language=\csname l@#1\endcsname
\fi
#2}}
\providecommand{\BIBdecl}{\relax}
\BIBdecl

\bibitem{fcPatent}
T.~L. Marzetta, O.~Blume, P.~Rulikowski, S.~Maier, A.~Pascht, and T.~Klein,
  ``Frequency division duplex ({FDD}) massive {MIMO} backhaul for repeater
  small cells,'' Patent US 2\,016\,134\,438 (A1), May 12, 2016.

\bibitem{1420804}
S.~Smith, ``Covariance, subspace, and intrinsic {C}ramér-{R}ao bounds,''
  \emph{IEEE Transactions on Signal Processing}, vol.~53, no.~5, pp.
  1610--1630, May 2005.

\bibitem{rcite}
\BIBentryALTinterwordspacing
{R Development Core Team}, \emph{R: A Language and Environment for Statistical
  Computing}, R Foundation for Statistical Computing, Vienna, Austria, 2008,
  {ISBN} 3-900051-07-0. [Online]. Available: \url{http://www.R-project.org}
\BIBentrySTDinterwordspacing

\bibitem{6897914}
A.~Checko, H.~L. Christiansen, Y.~Yan, L.~Scolari, G.~Kardaras, M.~S. Berger,
  and L.~Dittmann, ``Cloud {RAN} for mobile networks - a technology overview,''
  \emph{IEEE Communications Surveys Tutorials}, vol.~17, no.~1, pp. 405--426,
  Firstquarter 2015.

\bibitem{5595728}
T.~Marzetta, ``Noncooperative cellular wireless with unlimited numbers of base
  station antennas,'' \emph{IEEE Transactions on Wireless Communications},
  vol.~9, no.~11, pp. 3590--3600, November 2010.

\bibitem{1362898}
J.~N. Laneman, D.~N.~C. Tse, and G.~W. Wornell, ``Cooperative diversity in
  wireless networks: Efficient protocols and outage behavior,'' \emph{IEEE
  Transactions on Information Theory}, vol.~50, no.~12, pp. 3062--3080, Dec
  2004.

\bibitem{5161790}
T.~Riihonen, S.~Werner, R.~Wichman, and E.~Z. B., ``On the feasibility of
  full-duplex relaying in the presence of loop interference,'' in \emph{IEEE
  10th Workshop on Signal Processing Advances in Wireless Communications}, June
  2009, pp. 275--279.

\bibitem{1608543}
T.~Marzetta and B.~Hochwald, ``Fast transfer of channel state information in
  wireless systems,'' \emph{IEEE Transactions on Signal Processing}, vol.~54,
  no.~4, pp. 1268--1278, April 2006.

\bibitem{4527202}
L.~Withers, R.~Taylor, and D.~Warme, ``{E}cho-{MIMO}: A two-way channel
  training method for matched cooperative beamforming,'' \emph{IEEE
  Transactions on Signal Processing}, vol.~56, no.~9, pp. 4419--4432, Sept
  2008.

\bibitem{1658228}
D.~Samardzija and N.~Mandayam, ``Unquantized and uncoded channel state
  information feedback in multiple-antenna multiuser systems,'' \emph{IEEE
  Transactions on Communications}, vol.~54, no.~7, pp. 1335--1345, July 2006.

\bibitem{6777295}
J.~Choi, D.~Love, and P.~Bidigare, ``Downlink training techniques for {FDD}
  massive {MIMO} systems: Open-loop and closed-loop training with memory,''
  \emph{IEEE Journal of Selected Topics in Signal Processing}, vol.~8, no.~5,
  pp. 802--814, Oct 2014.

\bibitem{6816089}
X.~Rao and V.~Lau, ``Distributed compressive {CSIT} estimation and feedback for
  {FDD} multi-user massive {MIMO} systems,'' \emph{IEEE Transactions on Signal
  Processing}, vol.~62, no.~12, pp. 3261--3271, June 2014.

\bibitem{08123235}
S.~Wesemann, H.~Schlesinger, A.~Pascht, and O.~Blume, ``Measurement and
  characterization of the temporal behavior of fixed massive {MIMO} links,'' in
  \emph{21st International ITG Workshop on Smart Antennas}, March 2017.

\bibitem{LTEbook}
S.~Sesia, I.~Toufik, and M.~Baker, \emph{{LTE - The UMTS Long Term Evolution:
  From Theory to Practice}}.\hskip 1em plus 0.5em minus 0.4em\relax John Wiley
  \& Sons, 2009.

\bibitem{1053821}
T.~Goblick, ``Theoretical limitations on the transmission of data from analog
  sources,'' \emph{IEEE Transactions on Information Theory}, vol.~11, no.~4,
  pp. 558--567, Oct 1965.

\bibitem{1197846}
M.~Gastpar, B.~Rimoldi, and M.~Vetterli, ``To code, or not to code: lossy
  source-channel communication revisited,'' \emph{IEEE Transactions on
  Information Theory}, vol.~49, no.~5, pp. 1147--1158, May 2003.

\bibitem{4100513}
J.~Hahn, M.~Meurer, P.~W. Baier, and W.~Zirwas, ``Spread-spectrum based
  low-cost provision of downlink channel state information to the access points
  of {FDD} {OFDM} mobile radio systems,'' in \emph{IEEE Ninth International
  Symposium on Spread Spectrum Techniques and Applications}, Aug 2006, pp.
  10--15.

\bibitem{6399275}
M.~Taniguchi, H.~Murata, S.~Yoshida, K.~Yamamoto, D.~Umehara, S.~Denno, and
  M.~Morikura, ``Field experiments of linearly precoded multi-user {MIMO}
  system at 5{GH}z band,'' in \emph{2012 IEEE Vehicular Technology Conference
  (VTC Fall)}, Sept 2012, pp. 1--5.

\bibitem{edelman1998geometry}
A.~Edelman, T.~A. Arias, and S.~T. Smith, ``The geometry of algorithms with
  orthogonality constraints,'' \emph{SIAM journal on Matrix Analysis and
  Applications}, vol.~20, no.~2, pp. 303--353, 1998.

\bibitem{839985}
A.~Srivastava, ``A {Ba}yesian approach to geometric subspace estimation,''
  \emph{IEEE Transactions on Signal Processing}, vol.~48, no.~5, pp.
  1390--1400, May 2000.

\bibitem{6006540}
O.~Besson, N.~Dobigeon, and J.-Y. Tourneret, ``Minimum mean square distance
  estimation of a subspace,'' \emph{IEEE Transactions on Signal Processing},
  vol.~59, no.~12, pp. 5709--5720, Dec 2011.

\bibitem{1323251}
T.~Dahl, N.~Christophersen, and D.~Gesbert, ``Blind {MIMO} eigenmode
  transmission based on the algebraic power method,'' \emph{IEEE Transactions
  on Signal Processing}, vol.~52, no.~9, pp. 2424--2431, Sept 2004.

\bibitem{6288608}
H.~Q. Ngo and E.~G. Larsson, ``{EVD}-based channel estimation in multicell
  multiuser {MIMO} systems with very large antenna arrays,'' in \emph{IEEE
  International Conference on Acoustics, Speech and Signal Processing
  (ICASSP)}, March 2012, pp. 3249--3252.

\bibitem{7439748}
H.~Ghauch, T.~Kim, M.~Bengtsson, and M.~Skoglund, ``Subspace estimation and
  decomposition for large millimeter-wave {MIMO} systems,'' \emph{IEEE Journal
  of Selected Topics in Signal Processing}, vol.~10, no.~3, pp. 528--542, April
  2016.

\bibitem{6666355}
K.~Guo, Y.~Guo, and G.~Ascheid, ``On the performance of {EVD}-based channel
  estimations in {MU}-{M}assive-{MIMO} systems,'' in \emph{IEEE 24th Annual
  International Symposium on Personal, Indoor, and Mobile Radio Communications
  (PIMRC)}, Sept 2013, pp. 1376--1380.

\bibitem{6894457}
C.~Dehos, J.~L. González, A.~D. Domenico, D.~Kténas, and L.~Dussopt,
  ``Millimeter-wave access and backhauling: the solution to the exponential
  data traffic increase in 5{G} mobile communications systems?'' \emph{IEEE
  Communications Magazine}, vol.~52, no.~9, pp. 88--95, September 2014.

\bibitem{3924}
S.~Mumtaz, J.~Rodriguez, and L.~Dai, \emph{mmWave Massive MIMO, A Paradigm for
  5G}.\hskip 1em plus 0.5em minus 0.4em\relax Elsevier Science, 2016.

\bibitem{7063458}
Y.~Han, W.~Shin, and J.~Lee, ``Projection based feedback compression for {FDD}
  massive {MIMO} systems,'' in \emph{Globecom Workshops}, Dec 2014, pp.
  364--369.

\bibitem{6214417}
P.-H. Kuo, H.~Kung, and P.-A. Ting, ``Compressive sensing based channel
  feedback protocols for spatially-correlated massive antenna arrays,'' in
  \emph{IEEE Wireless Communications and Networking Conference}, April 2012,
  pp. 492--497.

\bibitem{6884026}
B.~Lee, J.~Choi, J.-Y. Seol, D.~Love, and B.~Shim, ``Antenna grouping based
  feedback reduction for {FDD}-based massive {MIMO} systems,'' in \emph{IEEE
  International Conference on Communications}, June 2014, pp. 4477--4482.

\bibitem{6881223}
Y.-G. Lim and C.~byoung Chae, ``Compressed channel feedback for correlated
  massive {MIMO} systems,'' in \emph{IEEE International Conference on
  Communications Workshops}, June 2014, pp. 360--364.

\bibitem{7302003}
W.~Shen, L.~Dai, B.~Shim, S.~Mumtaz, and Z.~Wang, ``Joint {CSIT} acquisition
  based on low-rank matrix completion for {FDD} massive {MIMO} systems,''
  \emph{IEEE Communications Letters}, vol.~19, no.~12, pp. 2178--2181, Dec
  2015.

\bibitem{7179337}
Z.~Gao, L.~Dai, W.~Dai, and Z.~Wang, ``Block compressive channel estimation and
  feedback for {FDD} massive {MIMO},'' in \emph{IEEE Conference on Computer
  Communications Workshops}, April 2015, pp. 49--50.

\bibitem{marzetta_larsson_yang_ngo_2016}
T.~L. Marzetta, E.~G. Larsson, H.~Yang, and H.~Q. Ngo, \emph{Fundamentals of
  Massive {MIMO}}.\hskip 1em plus 0.5em minus 0.4em\relax Cambridge University
  Press, 2016.

\bibitem{complBingham}
J.~T. Kent, ``The complex {B}ingham distribution and shape analysis,''
  \emph{Journal of the Royal Statistical Society, Series B (Methodological)},
  vol.~56, no.~2, 1994.

\bibitem{6086771}
N.~O'Donoughue and J.~M.~F. Moura, ``On the product of independent complex
  {G}aussians,'' \emph{IEEE Transactions on Signal Processing}, vol.~60, no.~3,
  pp. 1050--1063, March 2012.

\bibitem{svd}
G.~Golub and C.~{Van Loan}, \emph{Matrix Computations}, 2nd~ed.\hskip 1em plus
  0.5em minus 0.4em\relax Baltimore: Johns Hopkins University Press, 1989.

\bibitem{Marzetta99capacityof}
T.~L. Marzetta and B.~M. Hochwald, ``Capacity of a mobile multiple-antenna
  communication link in {R}ayleigh flat fading,'' \emph{IEEE Transactions on
  Information Theory}, vol.~45, pp. 139--157, 1999.

\bibitem{bertsekas1995nonlinear}
D.~Bertsekas, \emph{Nonlinear Programming}.\hskip 1em plus 0.5em minus
  0.4em\relax Athena Scientific, 1995.

\bibitem{6418045}
N.~Boumal, ``On intrinsic {C}ram\'{e}r-{R}ao bounds for {R}iemannian
  submanifolds and quotient manifolds,'' \emph{IEEE Transactions on Signal
  Processing}, vol.~61, no.~7, pp. 1809--1821, April 2013.

\end{thebibliography}

\end{document}